\documentclass[aps,prd,preprint,eqsecnum,floats,epsf,superscriptaddress,nofootinbib]{revtex4}
\usepackage{epsfig}
\usepackage{graphicx}
\usepackage{extarrows}
\usepackage{subfigure}

\def\ov{\overline}
\def\be{\begin{eqnarray}}
\def\en{\end{eqnarray}}
\def\non{\nonumber}

\def\la{\langle}
\def\ra{\rangle}
\def\A{{\cal A}}
\def\B{{\cal B}}

\def\lsim{ {\ \lower-1.2pt\vbox{\hbox{\rlap{$<$}\lower5pt\vbox{\hbox{$\sim$}
}}}\ } }
\def\gsim{ {\ \lower-1.2pt\vbox{\hbox{\rlap{$>$}\lower5pt\vbox{\hbox{$\sim$}
}}}\ } }

\begin{document}

\font\el=cmbx10 scaled \magstep2{\obeylines\hfill January, 2022}
\vskip 0.8 cm

\title{Hadronic three-body  $D$ decays mediated by scalar resonances}

\author{Hai-Yang Cheng}
\affiliation{Institute of Physics, Academia Sinica, Taipei, Taiwan 11529, ROC}

\author{Cheng-Wei Chiang}
\affiliation{Department of Physics, National Taiwan University, Taipei, Taiwan 10617, ROC}
\affiliation{Physics Division, National Center for Theoretical Sciences, Taipei, Taiwan 10617, ROC}

\author{Zhi-Qing Zhang}
\affiliation{Department of Physics, Henan University of Technology, Zhengzhou, Henan 450052, P.R. China}


\vskip 0.5cm
\begin{abstract}
\small
\vskip 0.5cm
We study the quasi-two-body $D\to SP$ decays and the three-body $D$ decays proceeding through intermediate scalar resonances, where $S$ and $P$ denote scalar and pseudoscalar mesons, respectively.
Our main results are:
(i) Certain external and internal $W$-emission diagrams with the emitted meson being a scalar meson are na{\"i}vely expected to vanish, but they actually receive contributions from vertex and hard spectator-scattering corrections beyond the factorization approximation.
(ii) For light scalars with masses below or close to 1~GeV, it is more sensible to study three-body decays directly and compare with experiment as the two-body branching fractions are either unavailable or subject to large finite-width effects of the scalar meson.
(iii) We consider the two-quark (scheme I) and four-quark (scheme II) descriptions of the light scalar mesons, and find the latter generally in better agreement with experiment. This is in line with recent BESIII measurements of semileptonic charm decays that prefer the tetraquark description of light scalars produced in charmed meson decays.
(iv) The topological amplitude approach fails here as the $D\to SP$ decay branching fractions cannot be reliably inferred from the measurements of three-body decays, mainly because the decay rates cannot be factorized into the topological amplitude squared and the phase space factor.
(v) The predicted rates for $D^0\to f_0 P, a_0 P$ are generally smaller than experimental data by one order of magnitude, presumably implying the significance of $W$-exchange amplitudes.
(vi) The $W$-annihilation amplitude is found to be very sizable in the $SP$ sector with $|A/T|_{SP}\sim 1/2$, contrary to its suppression in the $PP$ sector with $|A/T|_{PP}\sim 0.18$.
(vii) Finite-width effects are very important for the very broad $\sigma/f_0(500)$ and $\kappa/K_0^*(700)$ mesons. The experimental branching fractions $\B(D^+\to\sigma\pi^+)$ and $\B(D^+\to\bar\kappa^0 \pi^+)$ are thus corrected to be $(3.8\pm0.3)\times 10^{-3}$ and $(6.7^{+5.6}_{-4.5})\%$, respectively.

\end{abstract}

\maketitle
\small

%

\pagebreak

\section{Introduction}

In recent years many measurements of hadronic three-body and four-body decays of charmed mesons have been performed with Dalitz-plot amplitude analyses. Amplitudes describing $D$ meson decays into multibody final states are dominated by quasi-two-body processes, such as $D\to PP, VP, SP, AP$ and $TP$, where $P, V, S, A$ and $T$ denote pseudoscalar, vector, scalar, axial-vector and tensor mesons, respectively.
Among various $S$-, $P$- and $D$-wave intermediate resonances, the identification of the scalar mesons is rather difficult due to their broad widths and flat angular distributions.

Scalar mesons with masses lower than 2 GeV can be classified into two nonets: one nonet with masses below or close to 1 GeV, including $\sigma/f_0(500)$, $f_0(980)$, $a_0(980)$ and $\kappa/K_0^*(700)$; and the other nonet with masses above 1 GeV, including $a_0(1450)$, $K^*_0(1430)$, $f_0(1370)$, $f_0(1500)$ and $f_0(1710)$. Since the last three are all isosinglet scalars and only two of them can be accommodated in the quark model, implying a dominant scalar glueball content in one of the three isosinglets.

In this work, we shall study the quasi-two-body $D\to SP$ decays and the three-body $D$ decays proceeding through intermediate scalar resonances. In Tables~\ref{tab:DataSP} and \ref{tab:DataD0SP} we collect all the measured branching fractions of $D\to SP\to P_1P_2P$ decays available in the Particle Data Group (PDG) \cite{PDG}. It is clear that $f_0(980)$ and the $f_0$ family such as $f_0(1370)$,  $f_0(1500)$ and $f_0(1710)$ are observed in the three-body decays of $D^+, D^0$ and $D_s^+$, while $a_0(980)$ is seen exclusively in three-body $D^0$ decays (except for $D_s^+\to a_0^{+,0}\pi^{0,+}$). Contrary to $f_0(980)$ and $a_0(980)$ which are relatively easy to identify experimentally, the establishment of $\sigma$ and $\kappa$ is very difficult and controversial because their widths are so broad that their shapes are not clearly resonant. Nevertheless, their signals in three-body $D$ decays have been identified in $D^{+,0}\to\sigma \pi^{+,0}\to \pi^+\pi^-\pi^{+,0}$,  $D^+\to \bar\kappa^0 \pi^+\to K_S\pi^0\pi^+$ and $D^+\to \bar\kappa^0 K^+\to \pi^+K^-K^+$, respectively. Because of threshold and coupled-channel effects for $f_0(980)$ and $a_0(980)$ and the very broad widths for $\sigma$ and $\kappa$, it is no longer pertinent to use the conventional Breit-Wigner parametrization to describe their line shapes.

The $D \to SP$ decays and related three-body $D$ decays have been studied previously in Refs.~\cite{Kamal,Katoch,Buccella96,Fajfer,ChengSP,ElBennich,Boito,Cheng:SAT,Xie:2014tma, Dedonder:2014xpa,Loiseau:2016mdm,Dedonder:2021dmb}.
In the $D \to SP$ decays, the flavor diagram of each topology has two possibilities: one with the spectator quark in the charmed meson going to the pseudoscalar meson in the final state, and the other with the spectator quark ending up in the scalar meson.  We thus need two copies of each topological diagram to describe the decay processes.  Many of these decays have been observed in recent years through dedicated experiments and powerful Dalitz plot analysis of multi-body decays.  We will investigate whether an extraction of the sizes and relative strong phases of these amplitudes is possible.

One purpose of studying these decays is to check our understanding in the structures and properties of light even-parity scalar mesons.  Another goal is to learn the final-state interaction pattern in view of the rich resonance spectrum around the $D$ meson mass range.  Not only does this work update our previous study \cite{Cheng:SAT}, we also study the finite-width effect in the three-body decays mediated by the scalar mesons.  Such an effect is observed to be particularly important for decays involving $\sigma/f_0(500)$ and $\kappa/K_0^*(700)$ in the intermediate state because of their broad widths compared to their masses, respectively.  Therefore, one should be careful in the use of the narrow width approximation (NWA) to extract the $D \to SP$ two-body decays from the three-body decay rates.

This paper is organized as follows.  In Section~\ref{sec:status}, we review the current experimental status about how various $D \to SP$ decay branching fractions are extracted using the NWA from three-body decay rates.  In Section~\ref{sec:properties}, we discuss the two-quark $q\bar q$ and tetraquark pictures of the scalar nonet near or below 1~GeV along with the associated conundrums.  The decay constants and form factors required for subsequent numerical calculations are given in this section, too.  Section~\ref{sec:flavorapp} sets up the notation and formalism of flavor amplitude analysis, for both quark-antiquark and tetraquark pictures.  In Section~\ref{sec:facapp}, we take the factorization approach as an alternative toward analyzing these decays.  We also introduce line shapes for the scalar resonances when describing various three-body decays.  Section~\ref{sec:results} gives the results obtained based upon the approaches in the previous two sections for a comparison.  Section~\ref{sec:finitewidth} is devoted to the study of finite-width effect and how the NWA should be modified.  We summarize our findings in Section~\ref{sec:conclusions}.

\section{Experimental status \label{sec:status}}

It is known that three- and four-body decays of heavy mesons provide a rich laboratory for studying the intermediate-state resonances.  The Dalitz plot analysis of three-body or four-body decays of charmed mesons is a very useful technique for this purpose.  We are interested in $D\to SP$  decays followed by $S\to P_1P_2$.
The results of various experiments are summarized in Tables~\ref{tab:DataSP} and \ref{tab:DataD0SP}.  To extract the branching fraction for a $D\to SP$ decay, it is the usual practice to use the NWA:
 \be \label{eq:fact}
 \Gamma(D\to SP\to P_1P_2P)=\Gamma(D\to SP)_{\rm NWA} \B(S\to P_1P_2) ~.
 \en
Since this relation holds only in the $\Gamma_S\to 0$ limit, we put the subscript NWA to emphasize that $\B(D\to SP)$ thus obtained is under this limit.
Finite width effects will be discussed in Section~\ref{sec:finitewidth}.
For the branching fractions of two-body decays of scalar mesons, we shall use~\cite{PDG}
\be
 \B(a_0(980)\to\pi\eta)=0.850\pm0.017 ~,  &&
 \B(\sigma(500)\to\pi^+\pi^-)={2\over 3} ~,  \non  \\
 \B(f_0(1500)\to\pi\pi)=0.345\pm0.022 ~, &&
 \B(f_0(1710)\to K^+K^-)=0.292\pm0.027 ~, \\
 \B(K_0^{*0}(1430)\to K^+\pi^-)={2\over 3}(0.93\pm0.10) ~, &&
 \B(\kappa(700)\to K^+\pi^-)={2\over 3} ~,  \non
 \en
where we have applied the average of $\Gamma(a_0(980)\to K\ov K)/\Gamma(a_0(980)\to \pi\eta)=0.177\pm 0.024$ from PDG \cite{PDG} to extract the branching fraction of $a_0(980)\to \pi\eta$, assuming that its width is saturated by the $K\ov K$ and $\pi\eta$ modes.
For $f_0(1710)$ we have used the values of $\Gamma(f_0(1710)\to \pi\pi)/\Gamma(f_0(1710)\to K\ov K)=0.23\pm0.05$ and $\Gamma(f_0(1710)\to \eta\eta)/\Gamma(f_0(1710)\to K\ov K)=0.48\pm0.15$ from PDG together with the assumption of its width being saturated by $\pi\pi$, $K\ov K$ and $\eta\eta$ modes.
For $S=f_0(980)$ or $a_0(980)$, we are not able to extract the branching fractions of $D\to SP$ due to the lack of information of $\B(S\to P_1P_2)$ (except for $a_0(980)\to\pi\eta$), especially for $\B(S\to K\ov K)$ where the threshold effect must be taken into account.
For example, the NWA relation
\be
 \Gamma(D^+\to f_0(980) K^+\to K^+K^-K^+)=\Gamma(D^+\to f_0(980) K^+)\B(f_0(980)\to K^+K^-)
\en
cannot be applied  to extract the branching fraction of $D^+\to f_0(980) K^+$ due to the unknown ${\cal B}(f_0(980)\to K^+K^-)$.
Therefore, we will calculate the branching fractions of $\B(D\to SP\to P_1P_2P)$ directly and compare them with experiment (see Table~\ref{tab:DtoSP:theory} below).

\begin{table}[t!]
\caption{Experimental branching fractions of $(D^+,D_s^+)\to SP\to P_1P_2P$ decays.  For simplicity and convenience, we have dropped the mass identification for $\sigma(500)$,
$f_0(980)$, $a_0(980)$, $\kappa(700)$ and $K^*_0(1430)$.  Data are taken from Ref.~\cite{PDG} unless specified otherwise. We have applied the NWA given by Eq.~(\ref{eq:fact}) to extract the branching fractions of the two-body $D$ decay denoted by $\B(D\to SP)_{\rm NWA}$.
}
\label{tab:DataSP}
\footnotesize{
\begin{ruledtabular}
\begin{tabular}{l l l}
$\B(D\to SP; S\to P_1P_2)$  & $\B(D\to SP)_{\rm NWA}$ \\ \hline
 $\B(D^+\to f_0\pi^+; f_0\to\pi^+\pi^-)=(1.56\pm 0.33)\times 10^{-4}$ &
 \\
 $\B(D^+\to f_0(1370)\pi^+; f_0(1370)\to\pi^+\pi^-)=(8\pm4)\times 10^{-5}$ & $$ \\
 $\B(D^+\to f_0(1500)\pi^+; f_0(1500)\to\pi^+\pi^-)=(1.1\pm0.4)\times 10^{-4}$ & $\B(D^+\to f_0(1500)\pi^+)=(4.78\pm1.77)\times 10^{-4}$ \\
 $\B(D^+\to f_0(1710)\pi^+; f_0(1710)\to\pi^+\pi^-)<5\times 10^{-5}$ & $\B(D^+\to f_0(1710)\pi^+)<5.8\times 10^{-4}$  \\
 $\B(D^+\to f_0K^+; f_0\to \pi^+\pi^-)=(4.4\pm 2.6)\times 10^{-5}$ &
 \\
 $\B(D^+\to f_0K^+; f_0\to K^+K^-)=(1.23\pm 0.02)\times 10^{-5}$ \footnotemark[1] & \\
  $\B(D^+\to a_0(1450)^0\pi^+; a_0^0\to K^+K^-)=(4.5^{+7.0}_{-1.8})\times 10^{-4}$  &
 $$ \\
 $\B(D^+\to\sigma\pi^+; \sigma\to\pi^+\pi^-)=(1.38\pm0.12)\times 10^{-3}$ &
 $\B(D^+\to\sigma\pi^+)=({2.07\pm0.18})\times 10^{-3}$ \\
 $\B(D^+\to \bar\kappa^0 \pi^+; \bar \kappa^0\to K_S\pi^0)=(6^{+5}_{-4})\times 10^{-3}$ &
 $\B(D^+\to \bar\kappa^0 \pi^+)=({3.6^{+3.0}_{-2.4}})\%$ \\
 $\B(D^+\to \bar\kappa^0 K^+; \bar \kappa^0\to K^-\pi^+)=(6.8^{+3.5}_{-2.1})\times 10^{-4}$ &
 $\B(D^+\to \bar\kappa^0 K^+)=({1.0^{+0.5}_{-0.3}})\times 10^{-3}$ \\
 $\B(D^+\to\ov K_0^{*0}\pi^+; \ov K_0^{*0}\to K^-\pi^+)=(1.25\pm 0.06)\%$  &
 $\B(D^+\to\ov K_0^{*0}\pi^+)=(2.02\pm0.24)\%$ \\
 $\B(D^+\to\ov K_0^{*0}\pi^+; \ov K_0^{*0}\to K_S\pi^0)=(2.7\pm 0.9)\times 10^{-3}$  &
 $\B(D^+\to\ov K_0^{*0}\pi^+)=(1.74\pm0.61)\%$ \\
 $\B(D^+\to\ov K_0^{*0}K^+;\ov K_0^{*0}\to K^-\pi^+)=(1.82\pm 0.35)\times 10^{-3}$  &
 $D^+\to\ov K_0^{*0}K^+$  prohibited on-shell \\
\hline
 $\B(D_s^+\to f_0\pi^+; f_0\to K^+K^-)=(1.14\pm 0.31)\%$ &  \\
 $\B(D_s^+\to f_0\pi^+; f_0\to \pi^0\pi^0)=(2.1\pm 0.4)\times 10^{-3}$ \footnotemark[2] &  \\
 $\B(D_s^+\to S(980)\pi^+; S(980)\to K^+K^-)=(1.05\pm 0.07)\%$ \footnotemark[3], \footnotemark[4] &  \\
 $\B(D_s^+\to f_0(1370)\pi^+; f_0\to K^+K^-)=(7\pm 5)\times 10^{-4}$  &  \\
 $\B(D_s^+\to f_0(1370)\pi^+; f_0\to K^+K^-)=(7\pm 2)\times 10^{-4}$ \footnotemark[3]  &  \\
 $\B(D_s^+\to f_0(1370)\pi^+; f_0\to \pi^0\pi^0)=(1.3\pm 0.2)\times 10^{-3}$ \footnotemark[2] &  \\
 $\B(D_s^+\to f_0(1710)\pi^+; f_0\to K^+K^-)=(6.6\pm 2.8)\times 10^{-4}$  & $\B(D_s^+\to f_0(1710)\pi^+)=(2.26\pm0.98)\times 10^{-3}$ \\
 $\B(D_s^+\to f_0(1710)\pi^+; f_0\to K^+K^-)=(10\pm 4)\times 10^{-4}$ \footnotemark[3]  & $\B(D_s^+\to f_0(1710)\pi^+)=(3.42\pm1.40)\times 10^{-3}$ \\
 $\B(D_s^+\to a_0^{+,0}\pi^{0,+}; a_0^{+,0}\to \eta\pi^{+,0})=(1.46\pm0.27)\%$ \footnotemark[5] &
 $\B(D_s^+\to a_0^0\pi^+ +a_0^+\pi^0)=(1.72\pm0.32)\%$  \\
 $\B(D_s^+\to \ov K_0^{*0}K^+; \ov K_0^{*0}\to K^-\pi^+)=(1.8\pm0.4)\times 10^{-3}$ &
 $\B(D_s^+\to \ov K_0^{*0}K^+)=({ 2.9\pm0.7})\times 10^{-3}$  \\
 $\B(D_s^+\to \ov K_0^{*0}K^+; \ov K_0^{*0}\to K^-\pi^+)=(1.6\pm0.4)\times 10^{-3}$ \footnotemark[3] &
 $\B(D_s^+\to \ov K_0^{*0}K^+)=({ 2.6\pm0.7})\times 10^{-3}$  \\
 $\B(D_s^+\to K_0^{*0}\pi^+;  K_0^{*0}\to K^+\pi^-)=(5.0\pm3.5)\times 10^{-4}$ &
 $\B(D_s^+\to K_0^{*0}\pi^+)=({ 8.1\pm5.7})\times 10^{-4}$  \\
\end{tabular}
\footnotetext[1]{Assuming a fit fraction of 20\% for $D^+\to f_0(980)K^+$ in $D^+\to K^+K^-K^+$ decay \cite{LHCb:D+toKKK}.}
\footnotetext[2]{BESIII data taken from Ref. \cite{BESIII:Dspi+pi0pi0}.}
\footnotetext[3]{BESIII data taken from Ref. \cite{BESIII:DsKKpi}.}
\footnotetext[4]{$S(980)$ denotes both $f_0(980)$ and $a_0(980)$.}
\footnotetext[5]{The branching fraction is assigned to be $(2.2\pm0.4)\%$ by the PDG \cite{PDG}.
However, as pointed out in Ref. \cite{BESIII:Dstoa0pi}, the fraction of $D_s^+\to a_0(980)^{+(0)}\pi^{0(+)}, a_0(980)^{+(0)}\to \pi^{0(+)}\eta$ with respect to the total fraction of $D_s^+\to a_0(980)\pi,a_0(980)\to\pi\eta$ is evaluated to be 0.66. Consequently,
the branching fraction should be multiplied by a factor of 0.66 to become $(1.46\pm0.27)\%$.}
\end{ruledtabular}
}
\end{table}

\begin{table}[t]
\caption{Same as Table \ref{tab:DataSP} except for $D^0\to SP\to P_1P_2P$ decays.}
\label{tab:DataD0SP}
 \medskip
\footnotesize{
\begin{ruledtabular}
\begin{tabular}{l l l}
$\B(D\to SP; S\to P_1P_2)$  & $\B(D\to SP)_{\rm NWA}$ \\ \hline
 $\B(D^0\to f_0\pi^0; f_0\to \pi^+\pi^-)=(3.7\pm0.9)\times 10^{-5}$  &
 \\
 $\B(D^0\to f_0\pi^0; f_0\to K^+K^-)=(3.6\pm0.6)\times 10^{-4}$  &
  \\
 $\B(D^0\to f_0(1370)\pi^0; f_0\to \pi^+\pi^-)=(5.5\pm2.1)\times 10^{-5}$  &
  \\
 $\B(D^0\to f_0(1500)\pi^0; f_0\to \pi^+\pi^-)=(5.8\pm1.6)\times 10^{-5}$  &
 $\B(D^0\to f_0(1500)\pi^0)=(2.5\pm0.7)\times 10^{-4}$ \\
 $\B(D^0\to f_0(1710)\pi^0; f_0\to \pi^+\pi^-)=(4.6\pm1.6)\times 10^{-5}$  & $\B(D^0\to f_0(1710)\pi^0)=(3.7\pm1.4)\times 10^{-4}$
  \\

 $\B(D^0\to f_0\ov K^0; f_0\to \pi^+\pi^-)=(2.40^{+0.80}_{-0.46})\times 10^{-3}$  &
 \\
 $\B(D^0\to f_0\ov K^0; f_0\to K^+K^-)<1.8\times 10^{-4}$  &  \\
 $\B(D^0\to f_0(1370)\ov K^0; f_0\to \pi^+\pi^-)
 =(5.6^{+1.8}_{-2.6})\times 10^{-3}$  & \\
 $\B(D^0\to f_0(1370)\ov K^0; f_0\to K^+K^-)=(3.4\pm2.2)\times 10^{-4}$  &  \\
 $\B(D^0\to a_0^+K^-; a_0^+\to K^+\ov K^0)=(1.18\pm 0.36)\times 10^{-3}$ & \\
  $\B(D^0\to a_0^+K^-; a_0^+\to K^+\ov K^0)=(3.07\pm 0.84)\times 10^{-3}$ \footnotemark[1] & \\
 $\B(D^0\to a_0^-K^+; a_0^-\to K^-\ov K^0)<2.2\times 10^{-4}$ &  \\
 $\B(D^0\to a_0^0\ov K^0; a_0^0\to K^+K^-)=(5.8\pm 0.8)\times 10^{-3}$  &  \\
 $\B(D^0\to a_0^0\ov K^0; a_0^0\to K^+K^-)=(8.12\pm 1.80)\times 10^{-3}$ \footnotemark[1] & \\
 $\B(D^0\to a_0^0\ov K^0; a_0^0\to
 \eta\pi^0)=(2.40\pm0.56)\times 10^{-2}$ &  $\B(D^0\to a_0^0\ov K^0)=({ 2.83\pm0.66})\%$ \\
 $\B(D^0\to a_0^-\pi^+; a_0^-\to K^-K^0)=(2.6\pm 2.8)\times 10^{-4}$ &  \\
 $\B(D^0\to a_0^+\pi^-; a_0^+\to K^+\ov K^0)=(1.2\pm 0.8)\times 10^{-3}$ &  \\
 $\B(D^0\to a_0(1450)^-\pi^+; a_0^-\to K^-K^0)=(5.0\pm 4.0)\times 10^{-5}$ &  \\
%
 $\B(D^0\to a_0(1450)^+\pi^-; a_0^+\to K^+\ov K^0)=(6.4\pm 5.0)\times 10^{-5}$ &  \\
%
 $\B(D^0\to a_0(1450)^-K^+; a_0^-\to K^-K_S)< 0.6\times 10^{-3}$ \footnotemark[1] &  \\ 
 $\B(D^0\to \sigma\pi^0; \sigma\to \pi^+\pi^-)=(1.22\pm0.22)\times 10^{-4}$  &
 $\B(D^0\to \sigma\pi^0)=({1.8\pm0.3})\times 10^{-4}$ \\
 $\B(D^0\to K_0^{*-}\pi^+; K_0^{*-}\to \ov K^0\pi^-)=(5.34^{+0.80}_{-0.66})\times 10^{-3}$ &
 $\B(D^0\to K_0^{*-}\pi^+)=({ 8.6^{+1.6}_{-1.4}})\times 10^{-3}$ \\
 $\B(D^0\to K_0^{*-}\pi^+; K_0^{*-}\to K^-\pi^0)=(4.8\pm 2.2)\times 10^{-3}$ &
 $\B(D^0\to K_0^{*-}\pi^+)=(1.55\pm0.73)\%$ \\
 $\B(D^0\to \ov K_0^{*0}\pi^0; \ov K_0^{*0}\to K^-\pi^+)=
 (5.9^{+5.0}_{-1.6})\times 10^{-3}$ &
 $\B(D^0\to \ov K_0^{*0}\pi^0)=(9.5^{+8.1}_{ -2.8})\times 10^{-3}$  \\
  $\B(D^0\to K_0^{*+}\pi^-; K_0^{*+}\to K^0\pi^+)<2.8\times 10^{-5}$ &
 $\B(D^0\to K_0^{*+}\pi^-)<4.5\times 10^{-5}$ \\
\end{tabular}
\footnotetext[1]{BESIII data taken from Ref. \cite{BESIII:D0KKKS}.}
\end{ruledtabular}
}
\end{table}

\section{Physical properties of scalar mesons \label{sec:properties}}

It is known that the underlying structure of scalar mesons is not well established theoretically (see, {\it e.g.}, Refs.~\cite{Amsler,Close} for a review).  Scalar mesons with masses lower than 2~GeV can be classified into two nonets: one nonet with masses below or close to 1~GeV, including the isoscalars $f_0(500)$ (or $\sigma$), $f_0(980)$, the isodoublet $K_0^*(700)$ (or $\kappa$) and the isovector $a_0(980)$; and the other nonet with masses above 1~GeV, including $f_0(1370)$, $a_0(1450)$, $K^*_0(1430)$ and $f_0(1500)/f_0(1710)$.  If the scalar meson states below or near 1~GeV are identified as the conventional low-lying $0^+$ $q\bar q$ nonet, then the nonet states above 1~GeV could be excited $q\bar q$ states.

In the na{\"i}ve quark model, the flavor wave functions of the light scalars read
 \be
 && \sigma={1\over \sqrt{2}}(u\bar u+d\bar d) ~, \qquad\qquad~
 f_0= s\bar s ~, \non \\
 && a_0^0={1\over\sqrt{2}}(u\bar u-d\bar d) ~, \qquad\qquad a_0^+=u\bar d ~,
 \qquad a_0^-=d\bar u ~,  \\
 && \kappa^{+}=u\bar s ~, \qquad \kappa^{0}= d\bar s ~, \qquad~
 \bar \kappa^{0}=s\bar d ~,\qquad~ \kappa^{-}=s\bar u ~, \non
 \en
where an ideal mixing for $f_0$ and $\sigma$ is assumed as $f_0(980)$ is the heaviest one and $\sigma$ the lightest one in the light scalar nonet.  However, as summarized in Ref.~\cite{Cheng:SAT}, this simple picture encounters several serious problems:
\begin{enumerate}
\item
It is impossible to understand the mass degeneracy between $f_0(980)$ and $a_0(980)$, which is the so-called ``inverted spectrum problem.''
\item
The $P$-wave $0^+$ meson has one unit of orbital angular momentum which costs an energy around 500~MeV.  Hence, it should have a mass lying above rather than below 1~GeV.
\item
It is difficult to explain why $\sigma$ and $\kappa$ are much broader than $f_0(980)$ and $a_0(980)$ in width.
\item
The $\gamma\gamma$ widths of $a_0(980)$ and $f_0(980)$ are much smaller than na{\"i}vely expected for a $q\bar{q}$ state~\cite{bar85}.
\item
The radiative decay $\phi\to a_0(980)\gamma$, which cannot proceed if $a_0(980)$ is a pure $q\bar q$ state, can be nicely described by the four-quark nature of $a_0(980)$ \cite{Achasov:1987ts,Achasov:2003cn} or the kaon loop mechanism~\cite{Schechter06}.  Likewise, the observation of the radiative decay $\phi\to f_0(980)\gamma\to \pi\pi\gamma$ is also accounted for by the four-quark state of $f_0(980)$ \cite{Achasov:2003cn}.
\end{enumerate}

It turns out that these difficulties can be readily resolved in the tetraquark scenario where the four-quark flavor wave functions of light scalar mesons are symbolically given by \cite{Jaffe}
 \be \label{4quarkw.f.}
 && \sigma=u\bar u d\bar d ~, \qquad\qquad\qquad~~
 f_0= \frac{1}{\sqrt2} (u\bar u+d\bar d) s\bar s ~,  \non \\
 && a_0^0= \frac{1}{\sqrt2} (u\bar u-d\bar d) s\bar s ~,
 \qquad a_0^+=u\bar ds\bar s ~,
 \qquad a_0^-=d\bar us\bar s ~, \non \\
 && \kappa^+=u\bar sd\bar d ~, \qquad \kappa^0=d\bar su\bar u ~,
 \qquad \bar \kappa^0=s\bar du\bar u ~,
 \qquad \kappa^-=s\bar ud\bar d ~.
 \en
The four quarks $q^2\bar q^2$ can form an $S$-wave (rather than $P$-wave) $0^+$ meson without introducing one unit of orbital angular momentum.   This four-quark description explains naturally the inverted mass spectrum of the light nonet,
\footnote{However, it has been claimed recently in Ref.~\cite{Kuroda:2019jzm} that the inverse mass hierarchy can be realized in the $q\bar q$ picture through a $U(1)$ axial anomaly including explicit $SU(3)_F$ breaking. The anomaly term contributes to $a_0(980)$ with the strange quark mass and to $\kappa/K_0^*(700)$ with the up or down quark mass due to its flavor singlet nature. The current mass of the strange quark makes the $a_0$ meson heavier than the $\kappa$ meson.}
especially the mass degeneracy between $f_0(980)$ and $a_0(980)$,
and accounts for the broad widths of $\sigma$ and $\kappa$ while $f_0(980)$ and $a_0(980)$ are narrow because of the suppressed phase space for their decays to the kaon pairs.  Lattice calculations have confirmed that $a_0(1450)$ and $K_0^*(1430)$ are $q\bar q$ mesons, and suggested that $\sigma$, $\kappa$ and $a_0(980)$ are tetraquark mesonia \cite{Prelovsek,Mathur,Wakayama:scalar,Alexandrou:a0kappa,Alexandrou:a0}.

The inverted spectrum problem can also be alleviated in the scenario where the light scalars are dynamically generated from the meson-meson interaction, with the $f_0(980)$ and the $a_0(980)$ coupling strongly to the $K\ov K$ channel with isospin 0 and 1, respectively.  Indeed, the whole light scalar nonet appears naturally from properly unitarized chiral amplitudes for pseudoscalar-pseudoscalar scatterings~\cite{Oller:1997ng,Oller:1998hw}. Consequently, both $f_0(980)$ and $a_0(980)$ are good candidates of $K\ov K$ molecular states~\cite{Weinstein:1990gu}, while $\sigma$ and $\kappa$ can be considered as the bound states of $\pi\pi$ and $K\pi$, respectively.

In the na{\"i}ve two-quark model with ideal mixing for $f_0(980)$ and $\sigma(500)$, $f_0(980)$ is purely an $s\bar s$ state, while $\sigma(500)$ is an $n\bar n$ state with $n\bar n\equiv (\bar uu+\bar dd)/\sqrt{2}$.  However, there also exists some experimental evidence indicating that $f_0(980)$ is not a purely $s\bar s$ state. For example, the observation of $\Gamma(J/\psi\to f_0\omega)\approx {1\over 2}\Gamma(J/\psi\to f_0\phi)$ \cite{PDG} clearly shows the existence of the
non-strange and strange quark contents in $f_0(980)$.  Therefore, isoscalars $\sigma(500)$ and $f_0(980)$ must have a mixing
\be \label{eq:mixing}
 |f_0(980)\ra = |s\bar s\ra\cos\theta+|n\bar n\ra\sin\theta ~,
 \qquad |\sigma(500)\ra = -|s\bar s\ra\sin\theta+|n\bar n\ra\cos\theta ~.
 \en
Various mixing angle measurements have been discussed in the literature and summarized in Refs.~\cite{CCY,Fleischer:2011au}. A recent measurement of the upper limit on the branching fraction product $\B(\ov B^0\to J/\psi f_0(980))\times\B(f_0(980)\to \pi^+\pi^-)$ by LHCb leads to $|\theta|<30^\circ$~\cite{LHCb:theta}.
Likewise, in the four-quark scenario for light scalar mesons, one can also define a similar $f_0$-$\sigma$ mixing angle
  \be
 |f_0(980)\ra =|n\bar ns\bar s\ra\cos\phi
 +|u\bar u d\bar d\ra\sin\phi ~, \qquad
 |\sigma(500)\ra = -|n\bar ns \bar s\ra\sin\phi+|u\bar u d\bar d\ra\cos\phi ~.
 \en
It has been shown that $\phi=174.6^\circ$~\cite{Maiani}.

In reality, the light scalar mesons could have both two-quark and four-quark components.  Indeed, a real hadron in the QCD language should be described by a set of Fock states each of which has the same quantum number as the hadron.  For example,
\begin{eqnarray}\label{eq:fockexpansion}
|a^+(980)\rangle &=& \psi_{u\bar d}^{a_0} |u\bar d\rangle +
\psi_{u\bar dg}^{a_0} |u\bar d g\ra + \psi_{u\bar d s\bar s}^{a_0}
|u\bar d s \bar s\rangle+ \dots\,.
\end{eqnarray}
In the tetraquark model, $\psi_{u\bar d s\bar s}^{a_0} \gg \psi_{u\bar d}^{a_0}$, while it is the other way around in the two-quark model.  Although as far as the spectrum and decay are concerned, light scalars are predominately tetraquark states, their productions in heavy meson decays and in high energy hadron collisions are probably more sensitive to the two-quark component of the scalar mesons. For example, one may
wonder if the energetic $f_0(980)$ produced in $B$ decays is
dominated by the four-quark configuration as it requires to
pick up two energetic quark-antiquark pairs to form a fast moving
light tetraquark. Since the scalar meson production in charm decays is not energetic, it is possible that it has adequate time to form a tetraquark state. In principle, the two-quark and four-quark descriptions of the light scalars can be discriminated in the semileptonic charm decays.  For example, the ratio
\be
R={\B(D^+\to f_0\ell^+\nu)+\B(D^+ \to \sigma\ell^+\nu)
\over \B(D^+\to a_0^0\ell^+\nu)}
\en
is equal to 1 in the two-quark scenario and 3 in the four-quark model under the flavor SU(3) symmetry~\cite{Wang:2009azc}. Based on the BESIII measurements of $D^+\to a_0(980)^0e^+\nu_e$ \cite{BESIII:Dtoa0SL}, $D^+\to \sigma e^+\nu_e$ and the upper limit on $D^+\to f_0(980) e^+\nu_e$ \cite{BESIII:DtosigmaSL}, it follows that $R>2.7$ at 90\% confidence level. Hence, the BESIII results favor the SU(3) nonet tetraquark description of the $f_0(500)$, $f_0(980)$ and $a_0(980)$ produced in charmed meson decays.  A detailed analysis of BESIII and CLEO data on the decays $D^+\to \pi^+\pi^- e^+\nu_e$ and $_s^+\to \pi^+\pi^- e^+\nu_e$ in Ref. \cite{Achasov:2020qfx} also shows results in favor of the
four-quark nature of light scalar mesons $f_0(500)$ and $f_0(980)$.

The vector and scalar decay constants of the scalar meson are, respectively,  defined as
 \be \label{eq:Sdecayc}
 \la S(p)|\bar q_2\gamma_\mu q_1|0\ra=f_S p_\mu ~,
 \qquad \la S|\bar q_2q_1|0\ra=m_S\bar f_S ~.
 \en
The neutral scalar mesons $\sigma$, $f_0$ and $a_0^0$ cannot be produced via the vector current owing to charge conjugation invariance or conservation of vector current:
 \be
 f_{\sigma}=f_{f_0}=f_{a_0^0}=0 ~.
 \en
Applying the equation of motion to Eq.~(\ref{eq:Sdecayc}) yields
 \be \label{eq:EOM}
 \mu_Sf_S=\bar f_S ~, \qquad\quad{\rm with}~~\mu_S={m_S\over
 m_2(\mu)-m_1(\mu)} ~,
 \en
where $m_{2}$ and $m_{1}$ are the running current quark masses.  Therefore, the vector decay constant of the scalar meson $f_S$ vanishes in the SU(3) or isospin limit. The vector decay constants of $K^*_0(1430)$ and the charged $a_0(980)$ are non-vanishing, but they are suppressed due to the small mass difference between the constituent $s$ and $u$ quarks and between $d$ and $u$ quarks, respectively.     The scalar decay constants $\bar f_S$ have been computed in Ref.~\cite{CCY} within the framework of QCD sum rules.  For reader's conveneince, we list the scalar decay constants (in units of MeV) at $\mu=1$ GeV relevant to the present work
\be
&& \bar f_{f_0}=370\pm20, \qquad \bar f_{a_0}=365\pm20, \qquad \bar f_{\sigma}=350\pm20, \qquad \bar f_{\kappa}=340\pm20, \non \\
&& \bar f_{a_0(1450)}=460\pm50, \qquad f_{f_0(1500)}=490\pm50, \qquad \bar f_{K_0^*}=445\pm50.
\en
From Eq.~(\ref{eq:EOM}) we obtain (in units of MeV) \footnote{The vector decay constants of the scalar meson and its antiparticle are of opposite sign. For example, $f_{a_0(980)^+}=-1.3\,{\rm MeV}$ and $f_{a_0(980)^-}=1.3\,{\rm MeV}$.}
\be
|f_{a_0(980)^\pm}|=1.3\,, \qquad |f_{a_0(1450)^\pm}|=1.1\,, \qquad |f_\kappa|=45.5\,, \qquad  |f_{K^*_0(1430)}|=35.3\,.
\en
In short, the vector decay constants of scalar mesons are either zero or very small for non-strange scalar mesons.

Form factors for $D\to P,S$ transitions are defined by \cite{BSW}
 \be \label{eq:DSm.e.}
 \la P(p')|V_\mu|D(p)\ra &=& \left(P_\mu-{m_D^2-m_P^2\over q^2}\,q_ \mu\right)
F_1^{DP}(q^2)+{m_D^2-m_P^2\over q^2}q_\mu\,F_0^{DP}(q^2) ~, \non \\
\la S(p')|A_\mu|D(p)\ra &=& -i\Bigg[\left(P_\mu-{m_D^2-m_S^2\over
q^2}\,q_ \mu\right) F_1^{DS}(q^2)   +{m_D^2-m_S^2\over
q^2}q_\mu\,F_0^{DS}(q^2)\Bigg] ~,
 \en
where $P_\mu=(p+p')_\mu$ and $q_\mu=(p-p')_\mu$.  As shown in Ref.~\cite{CCH}, a factor of $(-i)$ is needed in the $D\to S$ transition in order for the $D\to S$ form factors to be positive.  This can also be checked from heavy quark symmetry consideration \cite{CCH}.

Throughout this paper, we use the 3-parameter parametrization
 \be \label{eq:FFpara}
 F(q^2)=\,{F(0)\over 1-a(q^2/m_D^2)+b(q^2/m_D^2)^2}
 \en
for $D\to S$ transitions. For hadronic $D\to SP$ decays, the relevant form factor is $F_0^{DS}(q^2)$. The parameters $F_0^{DS}(0)$, $a$ and $b$ for $D \to S$ transitions calculated in the covariant light-front quark model (CLFQM) \cite{CCH,Verma:2011yw}, covariant confined quark model (CCQM) \cite{Soni:2020sgn}, light-cone sum rules (LCSR) \cite{Shi:2017pgh,Cheng:2017fkw,Huang:2021owr} are exhibited in Table~\ref{tab:FFDtoS}. Note that the matrix element $\la S(p')|A_\mu|D(p)\ra$ is sometimes parametrized as
 \be
\la S(p')|A_\mu|D(p)\ra &=& -i\left[F_+^{DS}(q^2)P_\mu  + F_-^{DS}(q^2)q_\mu \right].
 \en
It is easily seen that
\be \label{eq:FFrel}
F_1(q^2)=F_+(q^2), \qquad F_0(q^2)={q^2\over m_D^2-m_S^2} F_-(q^2)+F_+(q^2)
~,
\en
and hence $F_1(0)=F_0(0)=F_+(0)$. It was argued in \cite{Huang:2021owr} that the relation $F_-(q^2)=-F_+(q^2)$ holds in the LCSR calculation.
In \cite{Soni:2020sgn}, the $D\to S$ transition form factors are defined by
\be
\la S(p)|A_\mu|D(p+q)\ra &=& -i\left[{F'}_+(q^2)p_\mu  + {F'}_-(q^2)q_\mu \right].
\en
They are related to $F_+(q^2)$ and $F_-(q^2)$ through the relation
\be
F'_+(q^2)=2F_+(q^2), \qquad F'_-(q^2)=F_+(q^2)+F_-(q^2).
\en

\begin{table}[t]
\caption{Form factors $F_0^{DS}(0)$ for $D, D_s\to f_0(980), a_0(980), a_0(1450)$ and $K_0^*(1430)$ transitions in various models. }
 \label{tab:FFDtoS}
  \medskip
\begin{ruledtabular}
\begin{tabular}{l c c c c c }
Transition & CLFQM & CCQM & LCSR(I) & LCSR(II) & LCSR(III) \\
 & \cite{CCH,Verma:2011yw} &  \cite{Soni:2020sgn} & \cite{Shi:2017pgh} & \cite{Cheng:2017fkw}
 & \cite{Huang:2021owr} \\
\hline
$D\to f_0(980)$ & $0.51^{+0.04}_{-0.05}$ \footnotemark[1] & $0.45\pm0.02$ & 0.321 &   \\
$D_s^+\to f_0(980)$ & $0.52^{+0.01}_{-0.01}$ \footnotemark[2] & $0.36\pm0.02$ &  & \\
$D\to a_0(980)$ \footnotemark[3] &   & $0.55\pm0.02$ &  & $0.88\pm0.13$ \footnotemark[4] & $0.85^{+0.10}_{-0.11}$ \\
$D\to a_0(1450)$ & $0.51^{+0.01}_{-0.02}$ & & & & $0.94^{+0.02}_{-0.03}$ \\
$D\to K_0^*(1430)$ & $0.47^{+0.02}_{-0.03}$ & \\
$D_s^+\to K_0^*(1430)$ & $0.55^{+0.02}_{-0.03}$  \\
\end{tabular}
\footnotetext[1]{For $D\to f_0^q$ transition.}
\footnotetext[2]{For $D_s^+\to f_0^s$ transition.}
\footnotetext[3]{It stands for either $D^0\to a_0(980)^-$ or $D^+\to a_0(980)^0$ transition.}
\footnotetext[4]{Use of the relation $F_+(0)=F'_+(0)/2$ has been made.}
\end{ruledtabular}
\end{table}

For the $q^2$ dependence of the form factors in various models, the parameters $a$ and $b$ are available in Refs. \cite{CCH,Verma:2011yw} and Ref. \cite{Shi:2017pgh} for CLFQM and LCSR(I), respectively. In CCQM and LCSR(II), one needs to apply Eq. (\ref{eq:FFrel}) to get the $q^2$ dependence of $F_0$. The form-factor $q^2$ dependence in the LCSR(III) calculation is shown in Fig. 3 of Ref. \cite{Huang:2021owr}.

BESIII has measured the branching fractions of both $D^0\to a_0(980)^-e^+\nu_e$ and $D^+\to a_0(980)^0e^+\nu_e$ \cite{BESIII:SLa0}. The theoretical calculations depend on the form factors $F_+(q^2)$ and $F_-(q^2)$ and their $q^2$ dependence (see e.g. Ref. \cite{Cheng:DmesonSL}). It turns out that the predicted branching fractions for $D\to a_0(980)e^+\nu_e$ in LCSR(II) \cite{Cheng:2017fkw} are too large by more than a factor of 2 compared to the BESIII experiment (see Table VI of  Ref. \cite{Huang:2021owr}). Hence, this model is disfavored.

\section{Diagrammatic amplitudes \label{sec:flavorapp}}

A least model-dependent analysis of heavy meson decays can be carried out in the so-called topological diagram approach. In this diagrammatic scenario, all two-body nonleptonic weak decays of heavy mesons can be expressed in terms of six distinct quark diagrams \cite{Chau,CC86,CC87}: $T$, the external $W$-emission tree diagram; $C$, the internal $W$-emission; $E$, the $W$-exchange; $A$, the $W$-annihilation; $H$, the horizontal $W$-loop; and $V$, the vertical $W$-loop.  The one-gluon exchange approximation of the $H$ graph is the so-called ``penguin diagram.''  These diagrams are classified according to the topologies of weak interactions with all strong interaction effects encoded.

The topological amplitudes for $D\to SP$ decays have been discussed in \cite{ChengSP,Cheng:SAT}. Just as $D\to V\!P$ decays,
one generally has two sets of distinct diagrams for each topology. For example, there are two
external $W$-emission and two internal $W$-emission diagrams, depending on whether the emitted particle is an even-party meson or an odd-parity one.  Following the convention in \cite{ChengSP,Cheng:SAT}, we shall denote the primed amplitudes $T'$ and $C'$ for the case when the emitted meson is a scalar one.  For the $W$-exchange and $W$-annihilation diagrams with the final state $q_1\bar q_2$, the primed amplitude denotes that the even-parity meson contains the quark $q_1$.
Since $K^*_0$, $a_0(1450)$ and the light scalars $\sigma,~\kappa,~f_0(980),~a_0(980)$ fall into two different SU(3) flavor nonets, in principle one cannot apply SU(3) symmetry to relate the topological amplitudes in $D^+\to f_0(980)\pi^+$ to, for example, those in $D^+\to \ov K^{*0}_0\pi^+$.

\begin{table}[!]
\caption{Topological amplitudes of various $D\to SP$ decays.  Schemes~I has $(\alpha, \beta) = (\sin\theta, \cos\theta)$, and scheme~II has $(\alpha , \beta) = (1, \sqrt2)$ for those modes with one $f_0$ and $(0,\sqrt2)$ for those modes with one $\sigma$.  In Scheme I, light scalar mesons $\sigma,~\kappa,~a_0(980)$ and $f_0(980)$ are described by the $q\bar q$  states, while $K^*_0$ and $a_0(1450)$ as excited $q\bar q$ states. In Scheme II, light scalars are tetraquark states, while $K^*_0$ and $a_0(1450)$ are ground-state $q\bar q$.  The $f_0-\sigma$ mixing angle $\theta$ in the two-quark model is defined in Eq. (\ref{eq:mixing}). The experimental branching fractions denoted by $\B_{\rm NWA}$ are taken from Tables \ref{tab:DataSP} and \ref{tab:DataD0SP}.
For simplicity, we do not consider the $f_0-\sigma$ mixing in the tetraquark model as its value is close to $\pi$ \cite{Maiani}.}
\label{tab:DSP}
\begin{ruledtabular}
\begin{tabular}{l l l }
Decay & Amplitude & $\B_{\rm NWA}$ \\
 \hline
 $D^+\to f_0\pi^+$ & $\frac{1}{\sqrt2}\alpha V_{cd}^*V_{ud}(T+C'+A+A') +\beta V_{cs}^*V_{us} C'$ & $$ \\
 \qquad $\to f_0K^+$ &$V_{cd}^*V_{us}\left[ {1\over\sqrt{2}}\alpha (T+A') + \beta A \right]$  \\
 \qquad $\to a_0^+\ov K^0$ & $V_{cs}^*V_{ud}(T'+C)$ &  \\
 \qquad $\to a_0^0\pi^+$ & $\frac{1}{\sqrt2} V_{cd}^*V_{ud}(-T-C'-A+A')$ & $$ \\
 \qquad $\to \sigma\pi^+$ & ${1\over\sqrt{2}}\beta V_{cd}^*V_{ud}(T+C'+A+A') - \alpha V_{cs}^*V_{us} C'$  & $(2.1\pm0.2)\times 10^{-3}$ \\
 \qquad $\to \bar\kappa^0\pi^+$ & $V_{cs}^*V_{ud}(T+C')$ & $(3.6^{+3.0}_{-2.4})\%$ \\
 \qquad $\to \bar\kappa^0K^+$ & $ V_{cs}^*V_{us}T + V_{cd}^*V_{ud}A$ & $(1.0^{+0.5}_{-0.3})\times 10^{-3}$ \\
 $D^0\to f_0\pi^0$ & ${1\over 2} \alpha V_{cd}^*V_{ud}(-C+C'-E-E') +{1\over\sqrt{2}} \beta V_{cs}^*V_{us} C'$ &  \\
  \quad~ $\to f_0\ov K^0$ & $V_{cs}^*V_{ud}[{1\over\sqrt{2}} \alpha (C+E) + \beta E']$ & \\
  \quad~ $\to a_0^+\pi^-$ & $V_{cd}^*V_{ud}(T'+E)$ & $$ \\
  \quad~ $\to a_0^-\pi^+$ & $V_{cd}^*V_{ud}(T+E')$ & $$ \\
  \quad~ $\to a_0^+K^-$ & $ V_{cs}^*V_{ud}(T'+E)$ & $$ \\
  \quad~ $\to a_0^0\ov K^0$ & $V_{cs}^*V_{ud}(C-E)/\sqrt{2}$ & $(2.83\pm0.66)\%$ \\
  \quad~ $\to a_0^-K^+$ & $ V_{cd}^*V_{us}(T+E')$ & \\
  \quad~ $\to \sigma\pi^0$ & ${1\over 2}V_{cd}^*V_{ud} \beta (-C+C'-E-E') -{1\over\sqrt{2}}\alpha V_{cs}^*V_{us} C'$ & $({ 1.8\pm0.3})\times 10^{-4}$ \\
 $D_s^+\to f_0\pi^+$ & $\frac{1}{\sqrt2} V_{cs}^*V_{ud} \left[ \sqrt2 \beta T+ \alpha (A+A') \right]$ &  \\
  \quad~ $\to f_0K^+$ & $V_{cs}^*V_{us}\left[\beta (T+C'+A) + {1\over \sqrt{2}}\alpha A' \right] +{1\over\sqrt{2}}V_{cd}^*V_{ud} \alpha C'$ \\
   \quad~ $\to a_0^0\pi^+$ & ${1\over \sqrt{2}}V_{cs}^*V_{ud}(-A+A')$ &
 $(0.86\pm0.23)\%$ \footnotemark[1]  \\
\hline
  $D^+\to a_0(1450)^{0}\pi^+$  & ${1\over\sqrt{2}}V_{cd}^*V_{ud}(-T-C'-A+A')$ & $$ \\
  \quad~ $\to \ov K_0^{*0}\pi^+$ & $V_{cs}^*V_{ud}(T+C')$  & $(1.98 \pm 0.22)\%$ \\
  \quad~ $\to \ov K_0^{*0}K^+$ & $V_{cs}^*V_{us}T + V_{cd}^*V_{ud}A$  &  prohibited \\
  $D^0\to a_0(1450)^{+}\pi^-$ &  $ V_{cd}^*V_{ud}(T'+E)$ &  \\
   \quad~ $\to a_0(1450)^{-}\pi^+$ &  $ V_{cd}^*V_{ud}(T+E')$ &   \\
   \quad~ $\to a_0(1450)^{-}K^+$ &  $ V_{cd}^*V_{us}(T+E')$ &   \\
   \quad~ $\to K_0^{*-}\pi^+$ & $V_{cs}^*V_{ud}(T+E')$ &   $(8.8\pm1.5)\times 10^{-3}$ \\
  \quad~ $\to \ov K_0^{*0}\pi^0$ &  ${1\over\sqrt{2}}V_{cs}^*V_{ud}(C'-E')$ & $(9.5^{+8.1}_{-2.8})\times 10^{-3}$ \\
   \quad~ $\to K_0^{*+}\pi^-$ & $V_{cd}^*V_{us}(T'+E)$ & $<4.5\times 10^{-5}$ \\
 $D_s^+ \to K_0^{*0}\pi^+$ & $V_{cd}^*V_{ud}\,T+V_{cs}V_{us}^*\,A$ & $(8.1\pm5.7)\times 10^{-4}$  \\
  \quad~ $\to \ov K_0^{*0}K^+$ & $V_{cs}^*V_{ud}(C'+A)$ & $(2.8\pm0.5)\times 10^{-3}$ \\
\end{tabular}
\footnotetext[1]{Since the decay amplitudes of $D_s^+\to a_0^+\pi^0$ and $D_s^+\to a_0^0\pi^+$
are the same except an overall negative sign, they have the same rates.}
\end{ruledtabular}
\end{table}

In Ref. \cite{Cheng:SAT} we have presented the topological amplitude decomposition in $D\to SP$ decays in two different schemes. In scheme I, light scalar mesons $\sigma, \kappa, a_0(980)$ and $f_0(980)$ are described by the ground-state $q\bar q$  states, while $K^*_0$ and $a_0(1450)$ as excited $q\bar q$ states. In scheme II, light scalars are tetraquark states, while $K^*_0$ and $a_0(1450)$ are ground-state $q\bar q$.
The topological amplitudes for $D\to SP$ decays are listed in Table~\ref{tab:DSP}.
The expressions of topological amplitudes are the same in both schemes I and II except for the channels involving $f_0$ and $\sigma$. For example,
\be \label{eq:AmpDtof0pi}
A(D^+\to f_0\pi^+) &=&  \left\{
\begin{array}{cl}
    {1\over\sqrt{2}}V_{cd}^*V_{ud}(T+C'+A+A')\sin\theta+V_{cs}^*V_{us}C'\cos\theta
      & \quad \mbox{Scheme~I} \ , \\
     {1\over\sqrt{2}}V_{cd}^*V_{ud}(T+C'+A+A')+\sqrt{2}V_{cs}^*V_{us}C'
      & \quad \mbox{Scheme~II} \ ,
    \end{array}\right.  \non \\
A(D^+\to \sigma\pi^+) &=&  \left\{
\begin{array}{cl}
    {1\over\sqrt{2}}V_{cd}^*V_{ud}(T+C'+A+A')\cos\theta-V_{cs}^*V_{us}C'\sin\theta
      & \quad \mbox{Scheme~I} \ , \\
     V_{cd}^*V_{ud}(T+C'+A+A')
      & \quad \mbox{Scheme~II} \ .
    \end{array}\right.
\en
In our numerical estimates, we will take $\theta = 30^\circ$, saturating the measured upper bound mentioned earlier.

In Table~\ref{tab:DSP} the upper part involves only light scalar mesons ($f_0$, $a_0$, $\sigma$, and $\kappa$), whereas the lower part involves the $a_0(1450)$ and $K_0^*(1430)$ mesons in the heavier nonet representation.  This division is made because the amplitudes of the same topology in these two groups have no {\it a priori} relations.  In each group we have 15 unknown parameters for the 8 topological amplitudes $T,C,E,A$ and $T',C',E',A'$. For neutral scalar mesons $\sigma,f_0$ and $a_0^0$, we cannot set $T'=C'=0$ even though their vector decay constants vanish. As will be discussed in
Sec. V.A, $T'$ and $C'$ do receive nonfactorizable contributions
through vertex and spectator-scattering corrections \cite{Cheng:2006,Cheng:2013}. Nevertheless, it is na{\"i}vely expected that, for example,  $|T'|\ll |T|$ and $|C'|\ll |C|$ for charged $a_0$.  However, as we shall see in Sec. V.C, a realistic calculation yields $|C'|>|C|$ instead.
At any rate, we have more theory parameters than observables (6 in the upper part and 5 in the lower part of the table), barring a fit.

Since the branching fractions of $f_0\to \pi\pi$ and $(f_0, a_0)\to K\ov K$ are unknown, many of the two-body decays in Table~\ref{tab:DSP} cannot be extracted from the data of three-body decays. Nevertheless, the strong couplings such as $g_{f_0\to \pi\pi}, g_{f_0\to K\bar K}, g_{a_0\to K\bar K}$ and $g_{a_0\to \eta\pi}$ have been inferred from a fit to the data. There are 17 available $D\to SP\to P_1P_2P_2$ modes, but there are only 14 data related to $D\to SP$ and we have 15 parameters to fit. Moreover, since we need to introduce appropriate energy-dependent line shapes for the scalar mesons, it is not conceivable to extract the topological amplitudes from three-body decays as the decay rate cannot be factorized into the topological amplitude squared and the phase space factor. We will come back to this point later.

It is interesting to notice that the current data already imply the importance of $W$-exchange and $W$-annihilation amplitudes. Consider the decays: $D^0\to a_0^+\pi^-\to K^+\ov K^0\pi^-$ and
$D^0\to a_0^-\pi^+\to K^-K^0\pi^+$ with the two-body decay amplitudes proportional to $(T'+E)$ and $(T+E')$, respectively (see Table~\ref{tab:DSP}).  If the $W$-exchange contributions are negligible, the former mode governed by the amplitude $T'$ is expected to have a rate smaller than the latter (cf. Table \ref{tab:DataD0SP}). Experimentally, it is the other way around. This is an indication that $E$ and $E'$ play some role.

\section{Factorization Approach \label{sec:facapp}}

The diagrammatic approach has been applied quite successfully to hadronic decays of charmed mesons into $PP$ and $V\!P$ final states \cite{RosnerPP08,RosnerVP,RosnerPP09,Cheng:Ddecay2010,%
Cheng:2012a,Cheng:2012b,Li:2012,Qin,Cheng:2016,Cheng:2021}.
When generalized to the decay modes involving a scalar meson in the final state, it appears that the current data are still insufficient for us to fully extract the information of all amplitudes.    Therefore, we take the na{\"i}ve factorization formalism as a complementary approach to estimate the rates of these decay modes.  In this framework, the $W$-exchange and -annihilation type of contributions will be neglected.

\subsection{Factorizable and nonfactorizable amplitudes}

The factorizable amplitudes for the $D\to SP$ decays read
 \begin{eqnarray} \label{eq:XDSP}
  X^{(D S, P)}
 &=& \langle P(q)| (V-A)_\mu|0\rangle \langle S(p)| (V-A)^\mu|D(p_D)\rangle, \non \\
  X^{(D P, S)}
 &=& \langle S(q)| (V-A)_\mu|0\rangle \langle P(p)| (V-A)^\mu|D(p_D)\rangle,
 \end{eqnarray}
and have the expressions
 \be \label{eq:XSP}
 X^{(DS, P)}
 =  -f_P(m_D^2-m_S^2) F_0^{DS}(q^2)\,,  \qquad
  X^{(D P, S)}=
 f_S (m_D^2-m_P^2) F_0^{DP}(q^2)\,,
 \end{eqnarray}
where use of Eqs.~(\ref{eq:Sdecayc}) and (\ref{eq:DSm.e.}) has been made. Hence,
\be \label{eq:SP_T,C}
T=- a_1(SP)f_P(m_D^2-m_S^2) F_0^{DS}(q^2), &\qquad& C=-a_2(SP)f_P(m_D^2-m_S^2) F_0^{DS}(q^2), \non \\
T'= a_1(PS)f_S (m_D^2-m_P^2) F_0^{DP}(q^2), &\qquad& C'=a_2(PS)f_S (m_D^2-m_P^2) F_0^{DP}(q^2).
\en
The primed amplitudes $T'$ and $C'$ vanish for the neutral scalar mesons such as $\sigma/f_0(500)$, $f_0(980)$ and $a_0(980)^0$ as they cannot be produced through the $(V-A)$ current; that is, $f_S=0$. Nevertheless, beyond the factorization approximation, contributions proportional to the scalar decay constant $\bar f_S$ of the scalar meson defined in Eq. (\ref{eq:Sdecayc})
can be produced from vertex and hard
spectator-scattering corrections. It has been shown in Refs. \cite{Cheng:2006,Cheng:2013} that the nonfactorizable  amplitudes can be recast to
\be \label{eq:S0P_T,C}
T'= a_1(PS)\bar f_S (m_D^2-m_P^2) F_0^{DP}(q^2), &\qquad& C'=a_2(PS)\bar f_S (m_D^2-m_P^2) F_0^{DP}(q^2),
\en
for $S=\sigma/f_0(500), f_0(980)$ and $a_0(980)^0$, etc., while the expressions of $T'$ and $C'$ given in Eq.~(\ref{eq:SP_T,C}) are valid for $S=a_0^\pm, \kappa/K^*_0(800)$ and $K_0^*(1430)$, etc.

\subsection{Flavor operators}

The flavor operators $a_i(M_1M_2)$ in Eqs. (\ref{eq:SP_T,C}) and (\ref{eq:S0P_T,C}) are basically the Wilson coefficients
in conjunction with short-distance nonfactorizable corrections such
as vertex corrections and hard spectator interactions. In general,
they have the expressions \cite{BBNS,BN}
\footnote{Notice that $a_1$ and $a_2$ do not receive contributions from penguin contractions.}
\be \label{eq:ai}
  a_1(M_1M_2) &=&
 \left(c_1+{c_2\over N_c}\right)N_1(M_2)  + {c_{2}\over N_c}\,{C_F\alpha_s\over
 4\pi}\Big[V_1(M_2)+{4\pi^2\over N_c}H_1(M_1M_2)\Big], \non \\
   a_2(M_1M_2) &=&
 \left(c_2+{c_1\over N_c}\right)N_2(M_2)  + {c_{1}\over N_c}\,{C_F\alpha_s\over
 4\pi}\Big[V_2(M_2)+{4\pi^2\over N_c}H_2(M_1M_2)\Big],
\en
where  $c_i$ are the Wilson coefficients,
$C_F=(N_c^2-1)/(2N_c)$ with $N_c=3$, $M_2$ is the emitted meson
and $M_1$ shares the same spectator quark with the $D$ meson. The
quantities $V_i(M_2)$ account for vertex corrections,
$H_i(M_1M_2)$ for hard spectator interactions with a hard gluon
exchange between the emitted meson and the spectator quark of the
$D$ meson.  The explicit expressions of $V_{1,2}(M)$ and $H_{1,2}(M_1M_2)$ in the QCD factorization approach are given in \cite{Cheng:2006}. The expression
of the quantities $N_i(M_2)$, which are relevant to the factorizable amplitudes, reads
\be \label{eq:Ni}
  N_i(P) = 1, \qquad N_i(S) = \begin{cases}
          0, \quad {\rm for}~S=\sigma, f_0, a_0^0, \\
          1, \quad {\rm else.}
          \end{cases}
\en
 Results for the flavor operators $a_i(M_1M_2)$ with $M_1M_2=SP$ and $PS$ are shown in Table \ref{tab:aiSP}. \footnote{Studies of $B\to SP$ decays in QCDF were presented in Refs. \cite{Cheng:2006,Cheng:2013}. Here We generalize these works to the $D\to SP$ decays and obtain the flavor operators given in Table \ref{tab:aiSP}.}

\begin{table}[t]
\caption{Numerical values of the flavor operators $a_{1,2}(M_1M_2)$ for $M_1M_2=SP$ and $PS$ at the scale $\mu=\ov m_c(\ov m_c)=1.3$ GeV, where use of $c_1(\mu)=1.33$ and $c_2(\mu)=-0.62$ has been made.}
\label{tab:aiSP}
\begin{center}
\begin{tabular}{ l c c | l r r} \hline \hline
$$ & ~~$f_0(500)\pi$~~ & ~~~$\pi f_0(500)$~~~ & ~~$$   & ~~~~$K_0^*(700)\pi$ ~~~~~ & ~~~$\pi K_0^*(700)$~~~~~~ \\
  \hline
 $a_1$ & ~~~$1.292+0.080i$~~~ & ~~$0.033-0.056i$~~ & ~~$a_1$~~ & $1.292+0.080i$ & $1.579-0.492i$ \\
 $a_2$ & $-0.527-0.172i$ & $-0.070+0.121i$ &  ~~$a_2$~~ & $-0.527-0.172i$ & $-1.147+0.930i$ \\
\hline\hline
 $$ & ~~$f_0(980)\pi$~~ & ~~~$\pi f_0(980)$~~~ & ~~$$~~  & ~~$f_0(980) K$~~~~~~ & $K f_0(980)$~~~~~~  \\
\hline
 $a_1$ & ~~~$1.292+0.080i$~~~ & ~~$0.033-0.056i$~~ & ~~$a_1$ & $1.295+0.075i$ & $0.033+0.075i$  \\
 $a_2$ & $-0.527-0.172i$ & $-0.070+0.121i$ &  ~~$a_2$ & $-0.533-0.162i$ & $-0.070+0.121i$ \\
\hline\hline
  $$ & ~~$a_0(980)^0\pi$~~ & ~~~$\pi a_0(980)^0$~~~ & ~~$$~~  & ~~$a_0(980)^0 K$~~~~~ & $K a_0(980)^0$~~~~ \\
  \hline
 $a_1$ & ~~~$1.292+0.080i$~~~ & ~~$0.037-0.066i$~~ & ~~$a_1$ & $1.295+0.075i$ & $0.037-0.066i$  \\
 $a_2$ & $-0.527-0.172i$ & $-0.080+0.141i$ &  ~~$a_2$ & $-0.533-0.162i$ & $-0.080+0.141i$ \\
\hline\hline
  $$ & ~~$a_0(980)^\pm\pi$~~ & ~~~$\pi a_0(980)^\pm$~~~ & ~~$$~~  & ~~$a_0(980)^\pm K$~~~~~ & $K a_0(980)^\pm$~~~ \\
  \hline
 $a_1$ & ~~~$1.292+0.080i$~~~ & ~~$\pm(-10.04+20.03i)$~~ & ~~$a_1$ & $1.295+0.075i$ & ~~~$\pm(-10.04+20.03i)$  \\
 $a_2$ & $-0.527-0.172i$ & ~~$\pm(23.89-43.14i)$ &  ~~$a_2$ & $-0.533-0.162i$ & ~~~$\pm(23.89-43.14i)$ \\
\hline\hline
   $$ & ~~$a_0(1450)\pi$~~ & ~~~$\pi a_0(1450)$~~~ & ~~$$~~  & ~~$K_0^*(1430)\pi$~~~~~ &
   $\pi K_0^*(1430)$~~~ \\
  \hline
 $a_1$ & ~~~$1.292+0.080i$~~~ & ~~$0.033-0.056i$~~ & ~~$a_1$~~ & $1.292+0.080i$ & ~~$1.692-0.544i$ \\
 $a_2$ & $-0.527-0.172i$ & $-0.071+0.108i$ &  ~~$a_2$~~ & $-0.527-0.172i$ & $-1.390+1.171i$ \\
\hline \hline
\end{tabular}
\end{center}
\end{table}

We see from Eqs.~(\ref{eq:ai}) and \eqref{eq:Ni} that the factorizable contributions to $a_1(PS)$ and $a_2(PS)$ vanish for $S=\sigma, f_0$ and $a_0^0$. Beyond the factorization approximation, nonfactorizable contributions proportional to the decay constant $\bar f_S$ can be produced from vertex and spectator-scattering corrections \cite{Cheng:2006,Cheng:2013}. Therefore, when the strong coupling $\alpha_s$ is turned off, the nonfactorizable contributions vanish accordingly. In short, the primed amplitudes $T'$ and $C'$  are factorizable  for $S=a_0^\pm, \kappa, K^*_0$, namely $\la S|J^\mu|0\ra\la P|J'_\mu|D\ra$, whereas they are nonfactorizable for $S=\sigma, f_0, a_0^0$.

Upon an inspection of Table \ref{tab:aiSP}, we see that (i) the flavor operators $a_i(PS)$ and $a_i(SP)$ are very different as the former does not receive factorizable contributions (i.e. $N_i(S)=0$), and (ii) while $a_1(SP)$ and $a_2(SP)$ are similar for any light and heavy scalar mesons, namely $a_1(SP)\approx 1.29\pm0.08i$ and $a_2(SP)\approx -0.53-0.17i$, $a_1(PS)$ and $a_2(PS)$ vary from neutral to the charged ones as shown in Table \ref{tab:aiPS}. One may wonder why the flavor operators $a_{1,2}(\pi a_0^\pm)$ are much greater than $a_{1,2}(\pi a_0^0)$. As noticed in Eqs. (\ref{eq:SP_T,C}) and (\ref{eq:S0P_T,C}), the nonfactorizable amplitudes are proportional to $a_{1,2}(\pi a_0^\pm)f_{a_0^\pm}$ for charged $a_0^\pm$ and to
$a_{1,2}(\pi a_0^0)\bar f_{a_0}$ for neutral $a_0^0$. Hence, $a_{1,2}(\pi a_0^\pm)/a_{1,2}(\pi a_0^0)=\bar f_{a_0}/f_{a_0^\pm}\gg 1$. We see from Table \ref{tab:aiPS} that $a_{1,2}(PS)$ become larger when the decay constants become smaller.

\begin{table}[t]
\caption{Same as Table \ref{tab:aiSP} except for the flavor operators $a_{1,2}(PS)$ with $P=\pi$. For neutral scalar mesons $\sigma,f_0,a_0^0$, the vector decay constant $f_S$ is replaced by the scalar decay constant $\bar f_S$. }
\label{tab:aiPS}
\begin{center}
\begin{tabular}{ c c c c} \hline \hline
 $S$ & ~~$f_S$ (MeV)~~ & ~~~$a_1(PS)$~~~ & ~~~$a_2(PS)$~  \\
\hline
 $\sigma,f_0,a_0^0$ & $350\sim 370$ & ~~$\sim 0.035-0.060i$~~ & ~~$\sim -0.075+0.130i$~~  \\
 $\bar \kappa$~~ &  $45.5$ & $1.58-0.49i$ & $-1.15+0.93i$  \\
 $\bar K_0^*$ & 35.3 & $1.69-0.54i$ & $-1.39+1.17i$  \\
 $a_0^-$ & 1.3 & $10-20i$ & $-24+43i$   \\
\hline \hline
\end{tabular}
\end{center}
\end{table}

\subsection{Implications}

Na{\"i}vely it is expected that $|T'(\pi^-a_0^+)|\ll |T(a_0^-\pi^+)|$ because $f_{\pi}\gg f_{a_0^+}$ and  $|C'(\pi^+\bar \kappa^0)|< |C(\pi^+f_0)|$ due to the fact that $f_\pi> f_\kappa$.  Although we are not able to extract the topological amplitudes of $D\to SP$ from the experimental data of three-body $D\to P_1P_2P_3$ decays, we can use the theoretical calculations to see their sizes and relative phases. From Eq. (\ref{eq:SP_T,C}) we have
\be
T(f_0\pi^+) &=& -a_1(f_0\pi)f_\pi(m_D^2-m_{f_0}^2)F_0^{Df_0}(m_\pi^2), \non \\
 C(f_0\pi^0) &=& -a_2(f_0\pi)f_\pi(m_D^2-m_{f_0}^2)F_0^{Df_0}(m_\pi^2),  \non \\
 T'(\pi^- a_0^+) &=& a_1(\pi a_0^+)f_{a_0^+}(m_D^2-m_{\pi}^2)F_0^{D\pi}(m_{a_0}^2), \\
 C'(\pi^0 f_0^0) &=& a_2(\pi f_0)\bar f_{f_0}(m_D^2-m_\pi^2)F_0^{D\pi}(m_{f_0}^2), \non \\
C'(\pi^+\bar\kappa^0) &=&  a_2(\pi \kappa)f_{\kappa}(m_D^2-m_\pi^2)F_0^{D\pi}(m_{\kappa}^2). \non
\en
Using the flavor operators given in Table \ref{tab:aiSP}, form factors $F^{DS}$ listed in Table \ref{tab:FFDtoS} and  $F^{DP}(q^2)$ evaluated in  the covariant confining quark model \cite{Ivanov:2019nqd}, we find numerically  (in units of $10^{-6}$ GeV),
\be
&& T(f_0\pi^+)=1.80\,e^{-i 186^\circ}, \quad~ C(f_0\pi^0)=0.77\, e^{-i 18^\circ}, \quad T'(\pi^-a_0^+)=0.55\, e^{i 117^\circ}, \non \\
&& C'(\pi^0 f_0)=0.99\, e^{i 120^\circ}, \quad~~ C'(\pi^+\bar\kappa^0)=1.26\, e^{i 141^\circ}.
\en
For heavier scalar mesons we find
\be
&& T(K_0^{*-}\pi^+)=0.70\,e^{-i 177^\circ}, \quad T'(\pi^-K_0^{*+})=1.29\, e^{-i 18^\circ}, \qquad
C'(\pi^0\bar K_0^{*0})=1.32\, e^{i 140^\circ}, \\
&& T(a_0(1450)^0\pi^+)=0.93\,e^{-i 177^\circ}, \quad T'(\pi^-a_0(1450)^+)=0.59\, e^{i 121^\circ}, \quad
C'(\pi^0a_0(1450)^{0})=1.21\, e^{i 123^\circ}. \non
\en
In the light scalar meson sector, we have $|T|>|T'|$ and  $|C|<|C'|$ rather than $|T|\gg|T'|$ and  $|C|>|C'|$. For scalar mesons in the higher nonet representation, we find $|T'|>|C'|>|T|$ with $|T|$ being suppressed as the mass term $(m_D^2-m_S^2)$ becomes smaller when $S$ becomes heavier.

\subsection{Flatt\'e line shape}
To describe three-body decays we need to introduce a line shape of the scalar resonance.
Normally we use the relativistic Breit-Wigner line shape to describe the scalar resonance contributions to three-body decays $D\to SP\to P_1P_2P$:
\be
T^{\rm BW}(s)={1 \over s-m_R^2+i m_R \Gamma_R(s)},
\en
with
\be
\Gamma_{R}(s)=\Gamma_{R}^0\left( {q\over q_0}\right)
{m_{R}\over \sqrt{s}},
\en
where $q=|\vec{p}_1|=|\vec{p}_2|$ is the c.m.~momentum in the rest frame of $R$, $q_0$ the value of $q$ when $s$ is equal to $m_R^2$.
However, this parametrization is not suitable to describe the decay of $f_0(980)$ or $a_0(980)$ into $K\ov K$ as $m(K^+)+m(K^-)=987.4$ MeV and $m(K^0)+m(\bar K^0)=995.2$ MeV are near threshold. In other words, one has to take the threshold effect into account. Since $f_0(980)$ couples strongly to the channel $K\ov K$ as well as to the channel $\pi\pi$, they can be described by a coupled channel formula, the so-called Flatt\'e line shape \cite{Flatte:1976xu}
\be \label{eq:Flattef0}
T^{\rm Flatte}_{f_0}(s)={1\over s-m_{f_0}^2+i\left[g_{f_0\to\pi\pi}^2\rho_{\pi\pi}(s)+g^2_{f_0\to K\bar K}\rho_{K\bar K}(s)\right]},
\en
with the phase space factor
\be
\rho_{ab} ={1\over 16\pi}\left(1-{(m_a+m_b)^2\over s}\right)^{1/2} \left(1-{(m_a-m_b)^2\over s}\right)^{1/2},
\en
so that
\be
\rho_{K\!\bar K}(s) &=& \rho_{K^+K^-}(s)+\rho_{K^0\bar K^0}(s)={1\over 16\pi}\left( \sqrt{1-(4m_{K^\pm}^2/ s)}+ \sqrt{1-(4m_{K^0}^2/ s)}\right), \non \\
\rho_{\pi\pi}(s) &=& \rho_{\pi^+\pi^-}(s)+{1\over 2}\rho_{\pi^0\pi^0}(s)={1\over 16\pi}\left( \sqrt{1-(4m_{\pi^\pm}^2/ s)}+{1\over 2} \sqrt{1-(4m_{\pi^0}^2/ s)}\right),
\en
and $\rho\to i\sqrt{-\rho^2}$ when below the threshold, i.e. $s<4m_K^2$ for $\rho_{K\bar K}$.
The dimensionful coupling constants in Eq. (\ref{eq:Flattef0}) are
\be
g_{f_0\to \pi\pi}\equiv g_{f_0\to \pi^+\pi^-}=\sqrt{2}g_{f_0\to \pi^0\pi^0}, \qquad
g_{f_0\to K\bar K}\equiv g_{f_0\to K^+K^-}=g_{f_0\to K^0\bar K^0}.
\en
Likewise, $a_0(980)$ couples strongly to $K\ov K$ and $\eta\pi$
\be
T^{\rm Flatte}_{a_0}(s)={1 \over s-m_{a_0}^2+i\left[g_{a_0\to\eta\pi}^2\rho_{\eta\pi}(s)+g^2_{a_0\to K\bar K}\rho_{K\bar K}(s)\right]}.
\en
with
\be
\rho_{\eta\pi}(s) &=& {1\over 16\pi}\left(1-{(m_\eta-m_\pi)^2\over s}\right)^{1/2} \left(1-{(m_\eta+m_\pi)^2\over s}\right)^{1/2}.
\en

It is important to check whether  $g_{f_0\to \pi\pi}$ and $g_{f_0,a_0\to K\bar K}$ can be interpreted as the strong couplings of $f_0$ to $\pi\pi$ and $K\ov K$, respectively. Using the formula
\be
\Gamma(f_0\to\pi^+\pi^-)={p_c\over 8\pi m_{f_0}^2}g_{f_0\to \pi^+\pi^-}^2,
\en
with $p_c$ being the c.m. momentum of the pion in the rest frame of $f_0$,
it is easily seen that the term $g_{f_0\to\pi\pi}^2\rho_{\pi\pi}(m_{f_0}^2)$ in Eq. (\ref{eq:Flattef0}) is identical to $m_{f_0}(\Gamma(f_0\to \pi^+\pi^-)+\Gamma(f_0\to \pi^0\pi^0))$. Therefore, we are sure that  $g_{f_0\to\pi\pi}$ is the strong coupling appearing in the matrix element $\la\pi^+\pi^-|f_0\ra$.
The strong couplings $g_{f_0,a_0\to K\bar K}$, $g_{f_0\to \pi\pi}$ and  $g_{a_0\to \eta\pi}$ have been extracted from fits to the experimental data. In this work we shall  use
\be \label{eq:couplings}
&& g_{f_0\to K\bar K}=(3.54\pm0.05)\,{\rm GeV}, \qquad~~
g_{a_0\to K\bar K}=(3.77\pm0.42)\,{\rm GeV},
\non \\ && g_{f_0\to\pi\pi}=(1.5\pm0.1)\,{\rm GeV}, \qquad\qquad~
g_{a_0\to \eta\pi}=(2.54\pm0.16)\,{\rm GeV},
\en
where the values of $g_{f_0\to K\bar K}$ and $g_{f_0\to \pi\pi}$ are taken from Ref. \cite{BESIII:D0KKKS},
dominated by the Dalitz plot analysis of $e^+e^-\to \pi^0\pi^0\gamma$ performed by KLOE \cite{KLOE:f0}. The couplings
$g_{a_0\to K\bar K}$ and $g_{a_0\to \pi\eta}$ are taken  from the analysis of the decay $D^0\to K_S^0K^+K^-$ by BESIII \cite{BESIII:D0KKKS}.
\footnote{From the amplitude analysis of the $\chi_{c1}\to \eta\pi^+\pi^-$ decay, BESIII obtained another set of couplings: $g_{a_0\to \eta\pi}=(4.14\pm0.02)\,{\rm GeV}$ and $g_{a_0\to K\bar K}=(3.91\pm0.02)\,{\rm GeV}$ \cite{BESIII:etapipi}. However, this set of couplings is not appealing for two reasons: (a) the large coupling constant $g_{a_0\to \eta\pi}$ will yield too large partial width $\Gamma_{\eta\pi}=222$ MeV, recalling that the total width of $a_0(980)$ lies in the range of 50 to 100 MeV \cite{PDG}, and (b) it is commonly believed that $a_0(980)$ couples more strongly to $K\ov K$  than to $\eta\pi$,
especially in the scenario in which $a_0(980)$ is a $K\ov K$ molecular state.}
Note the result for the coupling $g_{f_0\to\pi\pi}$ is consistent with the value of $1.33^{+0.29}_{-0.26}$ GeV extracted from Belle's measurement of the partial width of $f_0(980)\to\pi^+\pi^-$~\cite{Belle:f0}.

The partial widths can be inferred from the strong couplings listed in Eq.~(\ref{eq:couplings}) as
\be
\Gamma(f_0(980)\to \pi\pi)=(65.7\pm8.8)\,{\rm MeV}, \qquad
\Gamma(a_0(980)\to \eta\pi)=(85.2\pm10.7)\,{\rm MeV},
\en
though they are not directly measured.

\subsection{Line shape for $\sigma/f_0(500)$ \label{sec:line shape}}
As stressed in Ref. \cite{Pelaez:2015qba},  the scalar resonance $\sigma/f_0(500)$ is very broad and cannot be described by the usual Breit-Wigner line shape. Its partial wave amplitude does not resemble a Breit-Wigner shape with a clear peak and a simultaneous steep rise in the phase. The mass and width of the $\sigma$ resonance are identified from the associated pole position $\sqrt{s_\sigma}$ of the partial wave amplitude in the second Riemann sheet as $\sqrt{s_\sigma}=m_\sigma-i\Gamma_\sigma/2$~\cite{Pelaez:2015qba}. We shall follow the LHCb Collaboration~\cite{Aaij:3pi_2} to use a simple pole description
\be \label{eq:T sigma}
T_\sigma(s)={1\over s-s_\sigma}={1\over s-m_\sigma^2+\Gamma_\sigma^2(s)/4+im_\sigma\Gamma_\sigma(s)},
\en
with $\sqrt{s_\sigma}=m_\sigma-i\Gamma_\sigma/2$ and
\be
\Gamma_{\sigma}(s)=\Gamma_{\sigma}^0\left( {q\over q_0}\right)
{m_{\sigma}\over \sqrt{s}}.
\en
Using the isobar description of the $\pi^+\pi^-$ $S$-wave to fit the $B^+\to\pi^+\pi^-\pi^+$ decay data, the LHCb Collaboration found~\cite{Aaij:3pi_2}
\be \label{eq:sigmaMass}
\sqrt{s_\sigma}=(563\pm 10)-i(350\pm13)\,{\rm MeV},
\en
consistent with the PDG value of $\sqrt{s_\sigma}=(400-550)-i(200-350)\,{\rm MeV}$~\cite{PDG}.

In principle, we could also use a similar pole shape $T_\kappa(s)$
\be \label{eq:T kappa}
T_\kappa(s)={1\over s-s_\kappa}={1\over s-m_\kappa^2+\Gamma_\kappa^2(s)/4+im_\kappa\Gamma_\kappa(s)}.
\en
to describe the broad resonance $\kappa/K_0^*(700)$
and follow \cite{Pelaez:2020uiw} to use the latest result
\be \label{eq:kappaMass}
\sqrt{s_\kappa}=(648\pm 7)-i(280\pm16)\,{\rm MeV},
\en
determined from a dispersive data analysis. However, we find that this line shape together with the above pole mass and width will yield
a very huge and unreasonable result for the finite-width correction to $D^+\to\bar \kappa^0\pi^+$ (see Sec. VI.B below).
Hence, we will use the usual Breit-Wigner lineshape for $\kappa/K_0^*(700)$ and take the Breit-Wigner mass and width~\cite{PDG}
\be
m_{K_0^*(700)}^{\rm BW}=845\pm17\,{\rm MeV}, \qquad \Gamma_{K_0^*(700)}^{\rm BW}=468\pm30\,{\rm MeV}.
\en

\subsection{Three-body decays}

We take $D^+\to \sigma\pi^+\to \pi^+\pi^-\pi^+$  as an example to illustrate the calculation for the three-body rate.
The two-body decay amplitude for $D^+\to \sigma(m_{12})\pi^+$ with $m_{12}$ ($m_{12}^2\equiv (p_1+p_2)^2)$ being the invariant mass of the $\sigma$  is given by
\begin{align}
\begin{split}
A(D^+\to \sigma(m_{12})\pi^+)
=&
{G_F\over\sqrt{2}}V_{cd}^*V_{ud}\Big[ -a_1(\sigma\pi)f_\pi(m_D^2-s)F_0^{D\sigma}(m_\pi^2)
\\
&~~~ +a_2(\pi\sigma)\bar f_\sigma(m_D^2-m_\pi^2)F_0^{D\pi}(s)\Big].
\end{split}
\end{align}
Denoting $\A_\sigma\equiv A(D^+\to\sigma\pi^+\to \pi^+(p_1)\pi^-(p_2)\pi^+(p_3))$, we have
\be
\A_\sigma= g^{\sigma\to \pi^+\pi^-} F(s_{12},m_\sigma)\,T_\sigma(s_{12})A(D^+\to \sigma(s_{12}) \pi^+)+ (s_{12}\leftrightarrow s_{23}),
\en
where the $\sigma$ line shape $T_\sigma$ is given by Eq. (\ref{eq:T sigma}).
When $\sigma$ is off the mass shell, especially when $s_{12}$ is approaching the upper bound of $(m_D-m_\pi)^2$, it is necessary to account for the off-shell effect. For this purpose, we shall follow~\cite{Cheng:FSI} to introduce a form factor $F(s,m_R)$ parametrized as
\be \label{eq:FF for coupling}
F(s,m_R)=\left( {\Lambda^2+m_R^2 \over \Lambda^2+s}\right)^n,
\en
with the cutoff $\Lambda$ not far from the resonance,
\be
\Lambda=m_R+\beta\Lambda_{\rm QCD},
\en
where the parameter $\beta$ is expected to be of order unity. We shall use $n=1$, $\Lambda_{\rm QCD}=250$ MeV and $\beta=1.0\pm0.2$ in subsequent calculations.

The decay rate then reads
\begin{align}
\begin{split}
&
\Gamma(D^+\to \sigma\pi^+\to \pi^+\pi^-\pi^+)
\\
&= {1\over 2}\,{1\over(2\pi)^3 32 m_D^3}\int ds_{12}\,ds_{23} \Bigg\{ {|g^{\sigma\to \pi^+\pi^-}|^2 F(s_{12},m_\sigma)^2\over (s_{12}-m^2_{\sigma}+\Gamma_\sigma(s_{12})/4)^2+m_{\sigma}^2\Gamma_{\sigma}^2(s_{12})}
|A(D^+\to \sigma(m_{12})\pi^+)|^2
\\
&\qquad\qquad  +(s_{12}\leftrightarrow s_{23})+{\rm interference}
\Bigg\},
\end{split}
\end{align}
where the factor of ${1\over 2}$ accounts for the identical particle effect. The coupling constant $g^{\sigma\to \pi^+\pi^-}$ is determined by the relation
\be
\Gamma_{\sigma\to \pi^+\pi^-}={p_c\over 8\pi m_\sigma^2}g^2_{\sigma\to\pi^+\pi^-}.
\en

\begin{table}[!]
\caption{Branching fractions for various $D\to SP$ decays calculated in schemes~I and II. The upper part involves only light scalar mesons ($f_0$, $a_0$, $\sigma$, and $\kappa$), whereas the lower part involves the $a_0(1450)$ and $K_0^*(1430)$ mesons in the heavier nonet representation. The theoretical calculations are done in the factorization approach with both $W$-exchange and $W$-annihilation amplitudes being neglected. In scheme~I,  $K_0^*$ and $a_0(1450)$ are excited $q\bar q$ states. Hence, their predictions are not presented here. The $f_0-\sigma$ mixing angle $\theta$ is taken to be $30^\circ$ for scheme~I.
}
\label{tab:DSPfact}
\medskip
\footnotesize{
\begin{ruledtabular}
\begin{tabular}{l c c l}
Decay & Scheme I   & Scheme II & $\B_{\rm NWA}$ \\
 \hline
 $D^+\to \sigma\pi^+$ & $2.6\times 10^{-3}$ & $4.6\times 10^{-3}$ & $(2.1\pm0.2)\times 10^{-3}$ \\
 \qquad $\to \bar\kappa^0\pi^+$ & $6.1\%$ & $6.1\%$ & $(3.6^{+3.0}_{-2.4})\%$ \\
 \qquad $\to \bar\kappa^0K^+$ & $1.1\times 10^{-3}$ & $1.1\times 10^{-3}$ & $(1.0^{+0.5}_{-0.3})\times 10^{-3}$ \\
%
%
  $D^0\to a_0^0\ov K^0$  &  $4.2\times 10^{-3}$ & $4.2\times 10^{-3}$  & $(2.83\pm0.66)\%$ \\
  \quad~ $\to \sigma\pi^0$ & $3.2\times 10^{-5}$  & $7.8\times 10^{-5}$    & $({ 1.8\pm0.3})\times 10^{-4}$ \\
%
%
 $D_s^+\to a_0^0\pi^+$ &  0 &  0 &
 $(0.86\pm0.23)\%$   \\
\hline

  $D^+\to \ov K_0^{*0}\pi^+$ &   & $2.19\%$ & $(1.98 \pm 0.22)\%$ \\
  $D^0\to K_0^{*-}\pi^+$ &   & $2.1\times 10^{-3}$ &   $(8.8\pm1.5)\times 10^{-3}$ \\
  \quad~ $\to \ov K_0^{*0}\pi^0$ &   & $2.1\times 10^{-3}$ & $(9.5^{+8.1}_{-2.8})\times 10^{-3}$ \\
   \quad~ $\to K_0^{*+}\pi^-$ &  & $1.1\times 10^{-5}$ &   $<4.5\times 10^{-5}$ \\
 $D_s^+ \to K_0^{*0}\pi^+$ &   & $2.9\times 10^{-4}$ & $(8.1\pm5.7)\times 10^{-4}$  \\
  \quad~ $\to \ov K_0^{*0}K^+$ &  & $3.1\times 10^{-3}$ & $(2.8\pm0.5)\times 10^{-3}$ \\
\end{tabular}
\end{ruledtabular}}
\end{table}

\section{Results and Discussion \label{sec:results}}
In Tables \ref{tab:DSPfact} and \ref{tab:DtoSP:theory} we have calculated two-body $D\to SP$ and three-body $D\to SP\to P_1P_2P$ decays, respectively, in schemes I and II using the factorization approach with $W$-exchange and $W$-annihilation being neglected. We see from Table \ref{tab:DSP} that the decay modes $D^+\to a_0^+\ov K^0, \bar \kappa \pi^+$ and $\ov K_0^*\pi^+$ are free of $W$-annihilation contributions and they are ideal for testing the validity of the factorization approach. From Table \ref{tab:DtoSP:theory} it is evident that the calculated rates of $D^+\to\bar \kappa\pi^+\to K_S\pi^0\pi^+$ and $D^+\to \ov K^{*0}_0\pi^+\to (K\pi)^0\pi^+$ in scheme II are in agreement with experiment.
These modes are governed by the topologies $T+C'$ which interfere {\it constructively}. This is in contrast to the Cabibbo-favored (CF) $D^+\to \ov K^0\pi^+$ decay in the $P\!P$ sector where $T$ and $C$ contribute {\it destructively.}
For $(D^+, D_0, D_s^+)\to f_0 P; f_0\to P_1P_2$, predictions in scheme II are improved over that in scheme I and the discrepancies presumably arise from the $W$-exchange or $W$-annihilation amplitude. This implies that the tetraquark picture for light scalars works better than the quark-antiquark scenario.

Upon an inspection of Table \ref{tab:DSPfact}, the reader may wonder (i) why the branching fractions for $D\to (f_0,\sigma)P$ decays in scheme~II are always larger than that in scheme~I except for $D^0\to f_0\pi^0$, and (ii)
why the predicted branching fractions of $D^+\to \sigma\pi^+$ and $D^+\to \bar \kappa^0\pi^+$ are larger than experimental data, while the corresponding three-body decays agree with the measurements. For (i), we see from Table~IV and also Eq.~(\ref{eq:AmpDtof0pi}) that the $W$-emission decay amplitude involving $\sigma$ is suppressed by a factor of $\cos\theta/\sqrt{2}$ in scheme~I relative to that in scheme~II, while it is suppressed by a factor of $\sin\theta$ for the $W$-emission decay amplitude involving $f_0(980)$.  As a consequence our choice of $\theta = 30^\circ$, the branching fractions for $D\to (f_0,\sigma)P$ in scheme~II are always larger than scheme~I except for $D^0\to f_0\pi^0$. For (ii), it
has something to do with the finite-width effects of $\sigma$ and $\kappa$ as they are both very broad. We shall see in Sec.~\ref{sec:finitewidth} that the extraction of $\B(D\to SP)$ from the data is affected by the broad widths of both $\sigma$ and $\kappa$.

\begin{table}[!]
\caption{Branching fractions of various $D\to SP\to P_1P_2P$ decays calculated in schemes~I and II.  For simplicity and convenience, we have dropped the mass identification for
$f_0(980)$, $a_0(980)$ and $K^*_0(1430)$.  Data are taken from Tables \ref{tab:DataSP} and \ref{tab:DataD0SP}. In scheme~I,  $K_0^*$ and $a_0(1450)$ are excited $q\bar q$ states. Hence, their predictions are not presented here. The $f_0-\sigma$ mixing angle $\theta$ is taken to be $30^\circ$ for scheme~I.
}
\label{tab:DtoSP:theory}
\vskip 0.3cm
\footnotesize{
\begin{ruledtabular}
\begin{tabular}{l  c c c }
$D\to SP; S\to P_1P_2$  & Scheme I & Scheme II & Experiment \\ \hline
 $D^+\to f_0\pi^+; f_0\to\pi^+\pi^-$ & $7.6\times 10^{-5}$ & $2.2\times 10^{-4}$ & $(1.56\pm 0.33)\times 10^{-4}$
 \\
 $D^+\to f_0K^+; f_0\to \pi^+\pi^-$ & $3.6\times 10^{-7}$ & $1.2\times 10^{-5}$  & $(4.4\pm 2.6)\times 10^{-5}$
 \\
 $D^+\to f_0K^+; f_0\to K^+K^-$ & $2.5\times 10^{-7}$ & $8.4\times 10^{-6}$  & $(1.23\pm 0.02)\times 10^{-5}$  \\
 $D^+\to\sigma\pi^+; \sigma\to\pi^+\pi^-$ & $4.9\times 10^{-4}$ & $1.7\times 10^{-3}$  & $(1.38\pm0.12)\times 10^{-3}$  \\
 $D^+\to \bar\kappa^0 \pi^+; \bar \kappa^0\to K_S\pi^0$ & $5.4\times 10^{-3}$ & $5.4\times 10^{-3}$ & $(6^{+5}_{-4})\times 10^{-3}$ \\
 $D^+\to \bar\kappa^0 K^+; \bar \kappa^0\to K^-\pi^+$ & $3.7\times 10^{-4}$  & $3.7\times 10^{-4}$ & $(6.8^{+3.5}_{-2.1})\times 10^{-4}$ \\
%
 $D^0\to f_0\pi^0; f_0\to \pi^+\pi^-$ & $1.6\times 10^{-5}$ & $1.4\times 10^{-5}$  & $(3.7\pm0.9)\times 10^{-5}$
 \\
 $D^0\to f_0\pi^0; f_0\to K^+K^-$ & $1.1\times 10^{-5}$ & $8.8\times 10^{-6}$  & $(3.6\pm0.6)\times 10^{-4}$
  \\
 $D^0\to f_0\ov K^0; f_0\to \pi^+\pi^-$ & $9.0\times 10^{-6}$ & $3.0\times 10^{-4}$  & $(2.40^{+0.80}_{-0.46})\times 10^{-3}$
 \\
 $D^0\to f_0\ov K^0; f_0\to K^+K^-$ & $4.3\times 10^{-6}$ & $1.4\times 10^{-4}$ & $<1.8\times 10^{-4}$    \\
 $D^0\to a_0^+\pi^-; a_0^+\to K^+\ov K^0$ & $1.3\times 10^{-5}$  & $1.3\times 10^{-5}$  & $(1.2\pm 0.8)\times 10^{-3}$   \\
  $D^0\to a_0^-\pi^+; a_0^-\to K^-K^0$ & $2.9\times 10^{-4}$ & $2.9\times 10^{-4}$ & $(2.6\pm 2.8)\times 10^{-4}$   \\
 $D^0\to a_0^+K^-; a_0^+\to K^+\ov K^0$ & $2.2\times 10^{-4}$ & $2.2\times 10^{-4}$ & $(1.47\pm 0.33)\times 10^{-3}$  \\
 $D^0\to a_0^0\ov K^0; a_0^0\to K^+K^-$ & $3.4\times 10^{-4}$  & $3.4\times 10^{-4}$ & $(6.18\pm0.73)\times 10^{-3}$    \\
 $D^0\to a_0^0\ov K^0; a_0^0\to \eta\pi^0$ & $1.1\times 10^{-3}$ & $1.1\times 10^{-3}$ & $(2.40\pm0.56)\%$  \\
 $D^0\to a_0^-K^+; a_0^-\to K^-\ov K^0$ & $1.7\times 10^{-5}$ & $1.7\times 10^{-5}$ & $<2.2\times 10^{-4}$   \\
 $D^0\to \sigma\pi^0; \sigma\to \pi^+\pi^-$ & $2.2\times 10^{-5}$ & $2.0\times 10^{-4}$ & $(1.22\pm0.22)\times 10^{-4}$  \\
 $D_s^+\to f_0\pi^+; f_0\to K^+K^-$ & $2.5\times 10^{-3}$ & $5.1\times 10^{-3}$ & $(1.14\pm 0.31)\%$  \\
 $D_s^+\to a_0^{+,0}\pi^{0,+}; a_0^{+,0}\to \eta\pi^{+,0}$ & 0 & 0  & $(1.46\pm0.27)\%$ \\
\hline
 $D^+\to a_0(1450)^0\pi^+; a_0^0\to K^+K^-$ & $$  & $1.7\times 10^{-5}$ &
 $(4.5^{+7.0}_{-1.8})\times 10^{-4}$   \\
 $D^+\to\ov K_0^{*0}\pi^+; \ov K_0^{*0}\to K^-\pi^+$ &  & 1.38\% & $(1.25\pm 0.06)\%$ \\
 $D^+\to\ov K_0^{*0}\pi^+; \ov K_0^{*0}\to K_S\pi^0$ & $$ & $6.0\times 10^{-3}$ & $(5.4\pm 1.8)\times 10^{-3}$  \\
 $D^+\to\ov K_0^{*0}K^+;\ov K_0^{*0}\to K^-\pi^+$ & $$ & $7.6\times 10^{-5}$ & $(1.82\pm 0.35)\times 10^{-3}$  \\
 $D^0\to a_0(1450)^-\pi^+; a_0^-\to K^-K^0$ & $$ & $6.1\times 10^{-6}$ &
 $(5.0\pm 4.0)\times 10^{-5}$  \\
 $D^0\to a_0(1450)^+\pi^-; a_0^+\to K^+\ov K^0$ & $$ & $1.8\times 10^{-7}$ &
 $(6.4\pm 5.0)\times 10^{-5}$ \\
 $D^0\to a_0(1450)^-K^+; a_0^-\to K^-K_S$ & & & $< 0.6\times 10^{-3}$  \\
$D^0\to K_0^{*-}\pi^+; K_0^{*-}\to \ov K^0\pi^-$ & $$ & $8.3\times 10^{-4}$ & $(5.34^{+0.80}_{-0.66})\times 10^{-3}$  \\
 $D^0\to K_0^{*-}\pi^+; K_0^{*-}\to K^-\pi^0$ & $$ & $4.2\times 10^{-4}$ & $(4.8\pm 2.2)\times 10^{-3}$ \\
 $D^0\to \ov K_0^{*0}\pi^0; \ov K_0^{*0}\to K^-\pi^+$ & $$ & $9.6\times 10^{-4}$ & $(5.9^{+5.0}_{-1.6})\times 10^{-3}$   \\
 $D^0\to K_0^{*+}\pi^-; K_0^{*+}\to K^0\pi^+$ & $$ & $5.4\times 10^{-6}$ & $<2.8\times 10^{-5}$  \\
 $D_s^+\to K_0^{*0}\pi^+;  K_0^{*0}\to K^+\pi^-$ & $$ & $1.3\times 10^{-4}$ & $(5.0\pm3.5)\times 10^{-4}$   \\
 $D_s^+\to \ov K_0^{*0}K^+; \ov K_0^{*0}\to K^-\pi^+$ & $$ & $2.0\times 10^{-3}$  & $(1.7\pm0.3)\times 10^{-3}$   \\
\end{tabular}
\end{ruledtabular}
}
\end{table}

\subsection{$W$-annihilation amplitude \label{sec:annihilation}}

In the factorization calculations presented in Tables \ref{tab:DSPfact} and \ref{tab:DtoSP:theory}, we have neglected both $W$-exchange and $W$-annihilation amplitudes.
The $D_s^+\to a_0^+\pi^0+a_0^0\pi^+$ mode recently observed by BESIII \cite{BESIII:Dstoa0pi} proceeds only through the $W$-annihilation amplitudes. However, its branching fraction at a percent level is much larger than the other two $W$-annihilation channels $D_s^+\to \omega\pi^+$ and $\rho^0\pi^+$ whose branching fractions are $(1.92\pm0.30)\times 10^{-3}$ and $(1.9\pm1.2)\times 10^{-4}$, respectively \cite{PDG}. This implies that $|A(SP)| > |A(V\!P)|$. In other words, the $W$-annihilation amplitude plays a more significant role in the $SP$ sector than in the $V\!P$ one.

\begin{figure}[t]
\begin{center}
\vspace{10pt}
\includegraphics[width=100mm]{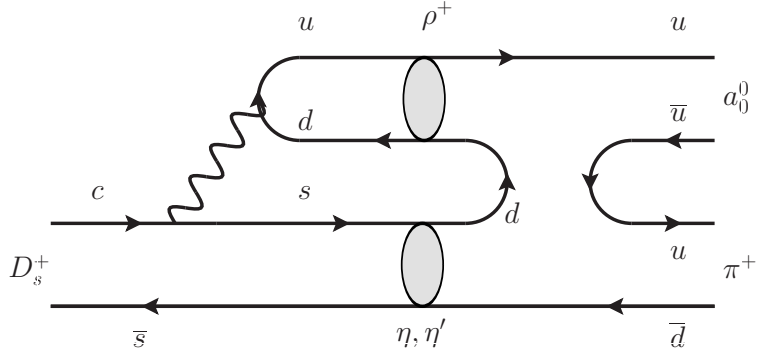}
\caption{Long-distance contributions to the $W$-annihilation amplitude of $D_s^+\to a_0^0\pi^+$ through final-state rescattering of $\rho\eta^{(')}\to a_0\pi$.
} \label{fig:Dsa0pi_FSI}
\end{center}
\end{figure}

\begin{figure}[t]
\begin{center}
\centering
\subfigure[]{
  \includegraphics[width=6.3cm]{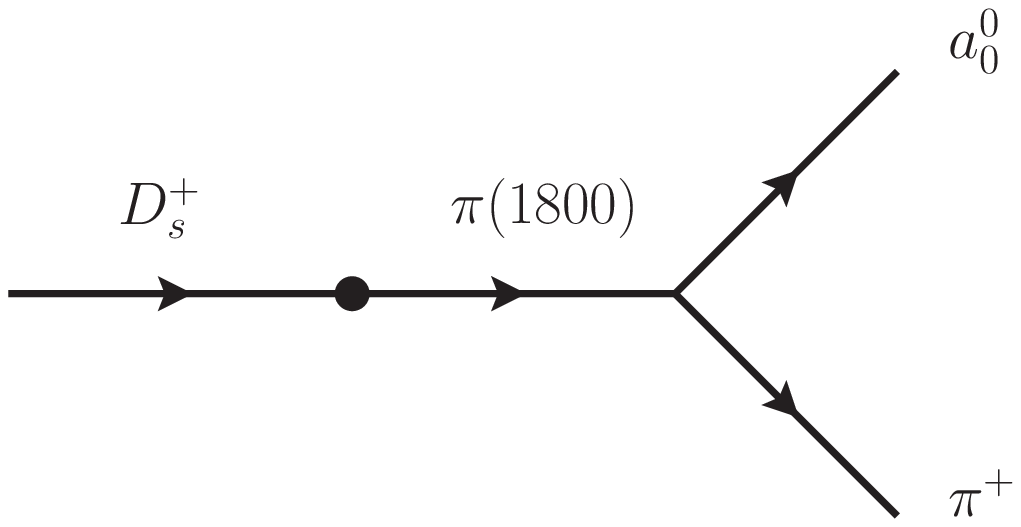}
  }
\hspace{0.5cm}
\subfigure[]{
  \includegraphics[width=6.7cm]{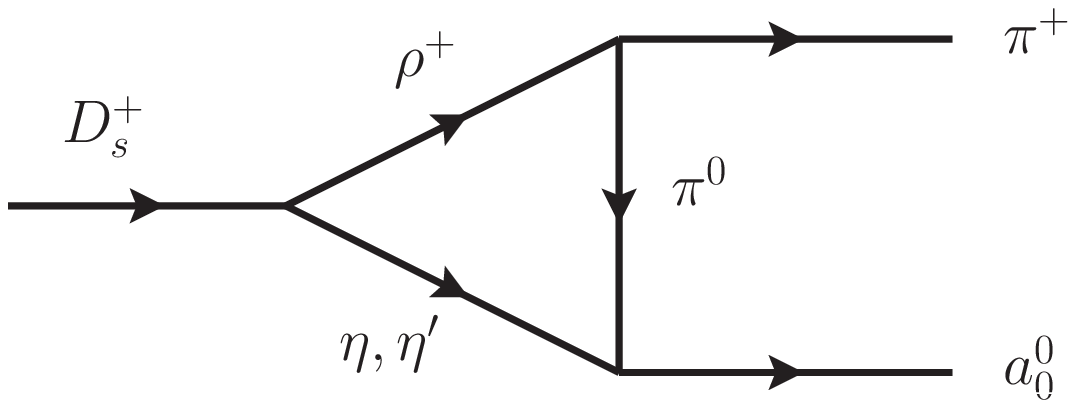}
}
\vspace{0.0cm}
\caption{Manifestation of Fig. \ref{fig:Dsa0pi_FSI} at the hadron level: (a) resonant contribution from the nearby resonance $\pi(1800)$ and (b) the triangle rescattering diagram.
} \label{fig:Dsa0pi}
\end{center}
\end{figure}

Consider the decay amplitude of $D_s^+\to a_0^0\pi^+$ and the $W$-annihilation contribution to $D_s^+\to f_0\pi^+$ (in scheme II)
\be
{\cal A}(D_s^+\to a_0^0\pi^+)={1\over\sqrt{2}}V_{cs}^*V_{ud}(-A+A'), \qquad
{\cal A}(D_s^+\to f_0\pi^+)_{\rm ann}={1\over\sqrt{2}}V_{cs}^*V_{ud}(A+A').
\en
Following the $G$-parity argument given in Ref. \cite{Cheng:Ddecay2010}, it is obvious that the direct $W$-annihilation process through $c\bar s\to W\to u\bar d$ is allowed in $D_s^+\to f_0\pi^+$ decay but not in $D_s^+\to a_0^0\pi^+$ decay as $G(u\bar d)=-$, $G(a_0\pi)=+$ and $G(f_0\pi)=-$.  This means that short-distance $W$-annihilation contributions respect the relation $A'=A$, contrary to the na{\"i}ve expectation.  Hence, one needs large long-distance $W$-annihilation which yields $A'=-A$.
Since $D_s^+\to\rho^+\eta$ has the largest branching fraction of $(8.9\pm0.8)\%$ among the CF $D_s^+\to VP$ decays \cite{PDG} , it is conceivable that
long-distance contribution from the weak decays $D_s^+\to \rho^+\eta$ followed by the resonantlike final-state rescattering of $\rho^+\eta\to a_0^0\pi^+$ (see Fig. \ref{fig:Dsa0pi_FSI}), which has the same topology as $W$-annihilation, may explain the large $W$-annihilation rate.
\footnote{The hadronic weak decays $D_s^+\to \rho^+\eta', \ov K^{*0}K^+$ and $\ov K^0K^{*+}$ followed by final-state rescattering will also contribute to $D_s^+\to a_0^0\pi^+$.}
It is customary to evaluate the final-state rescattering contribution, Fig. \ref{fig:Dsa0pi_FSI}, at the hadron level manifested in Fig. \ref{fig:Dsa0pi}. One of the diagrams, namely, the triangle graph in Fig. \ref{fig:Dsa0pi}(b) has been evaluated recently in \cite{Hsiao:a0,Ling:a0}. It yields a major contribution to $D_s^+\to a_0^0\pi^+$ owing to the large coupling constants for $\rho^+\to\pi^+\pi^0$ and $a_0^0\to \pi^0\eta$. The graph in Fig. \ref{fig:Dsa0pi}(a) shows the resonant final-state interactions manifested by the nearby resonance
$\pi(1800)$ whose strong decay to $a_0\pi$ has been seen experimentally \cite{PDG}. However, we are not able to have a quantitative statement owing to the lack of information on its partial width.

Assuming $A'\approx -A$, the annihilation amplitude extracted from the data of $D_s^+\to a_0^+\pi^0+a_0^0\pi^+$ is (in units of $10^{-6}$ GeV),
\be
|A|=0.91\pm0.12 \,.
\en
Hence, the annihilation amplitude is very sizable in the $SP$ sector, $|A/T|_{SP}\sim 1/2$, contrary to its suppression  $|A/T|_{PP}\sim 0.18$ in the $P\!P$ sector \cite{Cheng:2019ggx} and $|A_V/T_P|_{VP}\sim 0.07$ in the $VP$ sector \cite{Cheng:2021yrn}.

\subsection{Finite Width Effects \label{sec:finitewidth}}

The finite-width effect is accounted for by the quantity $\eta_R$ defined by \cite{Cheng:2020mna,Cheng:2020iwk}
\be \label{eq:eta}
\eta_{_R}\equiv \frac{\Gamma(D\to RP_3\to P_1P_2P_3)_{\Gamma_R\to 0}}{\Gamma(D\to RP_3\to P_1P_2P_3)}=\frac{\Gamma(D\to RP_3)\B(R\to P_1P_2)}{\Gamma(D\to RP_3\to P_1P_2P_3)}=1+\delta
~,
\en
so that the deviation of $\eta_{_R}$ from unity measures the degree of departure from the NWA when the resonance width is finite.  It is na{\"i}vely expected that the correction $\delta$ will be of order $\Gamma_R/m_R$. It is calculable theoretically but  depends on the line shape of the resonance and the approach of describing weak hadronic decays such as QCD factorization and perturbative QCD.

Using the branching fractions of two-body and three-body $D$ decays calculated in Tables \ref{tab:DSPfact} and \ref{tab:DtoSP:theory}, respectively, in scheme II,
the resultant $\eta_R$ parameters for scalar resonances $\sigma, \kappa$ and $K_0^*$ produced in the three-body $D$ decays are summarized in Table~\ref{tab:eta}. We only consider the $D^+$ decays as the three-body modes listed in Table~\ref{tab:eta} are not contaminated by the $W$-annihilation amplitude and hence the calculated finite width effects are more trustworth. We have also checked explicitly that $\eta_R\to 1$ in the narrow width limit as it should be. The $\eta_R$ parameters for various resonances produced in the three-body $B$ decays have been evaluated in
\cite{Cheng:2020mna,Cheng:2020iwk}. Our results for $\eta_R$'s in Table~\ref{tab:eta} have similar features as the values $\eta_{\sigma/f_0(500)}=2.15\pm0.05$ and $\eta_{K_0^*(1430)}=0.83\pm0.04$ obtained in $B$ decays.

\begin{table}[t]
\caption{A summary of the $\eta_R$ parameter for scalar resonances produced in the three-body $D$ decays. The mass and width of $\sigma/f_0(500)$ are taken from Eq. (\ref{eq:sigmaMass}).
}
\vskip 0.15cm
\label{tab:eta}
\footnotesize{
\begin{ruledtabular}
\begin{tabular}{ l l c c c l }
 Resonance~~~ & ~$D\to Rh_3\to h_1h_2h_3$ ~~~ & ~$\Gamma_R$ (MeV)~\cite{PDG}~~ & ~$m_R$ (MeV)~\cite{PDG} &  $\Gamma_R/m_R$ & ~~~$\eta_R$ \\
\hline
$\sigma/f_0(500)$ & $D^+\to \sigma\pi^+\to \pi^+\pi^-\pi^+$ & ~$700\pm26$~~ & ~$563\pm10$~~ & $1.243\pm0.051$ & ~~1.850 \\
$\kappa/K_0^*(700)$ & $D^+\to \bar\kappa^0\pi^+\to K_S^0\pi^0\pi^+$ & ~$468\pm30$~~ & $845\pm17$ & $0.554\pm0.037$ & ~~1.873 \\
$K_0^*(1430)$ & $D^+\to \ov K_0^{*0}\pi^+\to K^-\pi^+\pi^+$ & ~$270\pm80$~~ & ~$1425\pm50$~~ & $0.19\pm0.06$ & ~~0.985 \\
\end{tabular}
\end{ruledtabular} }
\end{table}
Note  that {\it a priori} we do not know if the deviation of $\eta_R$ from unity is positive or negative. In general, it depends on the line shape, mass and width of the resonance.  As alluded to above, the mass and width have a more dominant effect than the line shape in the case of $\kappa(700)$. As another example, we found in Ref.~[83] that $\eta_\rho>1$ for the Breit-Wigner line shape and $\eta_\rho<1$ when the Gounaris-Sakurai model \cite{Gounaris:1968mw} is used to describe the line shape of the broad $\rho(770)$ resonance. To our knowledge, there is no good argument favoring one line shape over the other.  Therefore, $\eta_{K_0^*(1430)}=0.985<1$, for example, is the result of our particular line shape choice.

When the resonance is sufficiently broad, it is necessary to take into account the finite-width effects characterized by the parameter $\eta_R$. Explicitly \cite{Cheng:2020mna,Cheng:2020iwk},
\be
\B(D\to RP)=\eta_R\B(D\to RP)_{\rm NWA}=\eta_R{\B(D\to RP_3\to P_1P_2P_3)_{\rm expt}\over
\B(R\to P_1P_2)_{\rm expt}}
~,
\en
Therefore, the experimental branching fractions $\B(D\to R P)_{\rm NWA}$ for $D^+\to\sigma \pi^+, \bar\kappa^0 \pi^+$ and $\ov K_0^{*0} \pi^+$ decays in Tables \ref{tab:DataSP} and \ref{tab:DSPfact} should have the following corrections:
\be
\B(D^+\to\sigma \pi^+): && (2.1\pm0.2)\times 10^{-3}\to  (3.8\pm0.3)\times 10^{-3},  \non \\
\B(D^+\to\bar\kappa^0 \pi^+): && (3.6^{+3.0}_{-2.4})\%  \qquad\quad  ~\to   (6.7^{+5.6}_{-4.5})\%,   \\
\B(D^+\to\ov K_0^{*0} \pi^+): && (1.98\pm0.22)\%  \quad ~~\to (1.94\pm0.22)\%. \non
\en
From Table \ref{tab:DSPfact}, it is evident that the agreement between theory and experiment is substantially improved for $D^+\to\sigma \pi^+$ and $D^+\to \bar\kappa^0 \pi^+$.

If we employ the pole mass and width, $m_\kappa=648\pm7$ MeV and $\Gamma_\kappa=560\pm 32$ MeV, respectively, for $\kappa/K_0^*(700)$ and the pole line shape given in Eq. (\ref{eq:T kappa}), we will be led to the results $\B(D^+\to \bar \kappa^0\pi^+)=8.10\%$, $\B(D^+\to \bar \kappa^0\pi^+\to K_S^0\pi^0\pi^+)=1.62\times 10^{-3}$ and $\eta_\kappa=8.34$. This implies that the finite-width correction will be unreasonably too large and thus unlikely, as alluded to at the end of Sec.~\ref{sec:line shape}.  However, if the Breit-Wigner mass and width are used instead, we get $\eta_\kappa=1.92$ for pole line shape, which is a more reasonable result. This implies that in this case, it is the mass and width rather than the line shape that governs the finite-width correction.

For the case of $f_0(500)$, one may wonder what the correction will be if the Breit-Wigner line shape is used. According to PDG \cite{PDG}, the Breit-Wigner mass and width of $f_0(500)$ lie in the wide ranges of 400-800~MeV and 100-800~MeV, respectively. As a result, it is quite difficult to pin down a specific set of parameters and thereby determine the finite-width correction. On the contrary, LHCb has determined its pole mass and width with reasonable accuracy using the pole line shape [see Eq. (\ref{eq:sigmaMass})]. It turns out that the pole mass and width fall within the above allowed ranges of the Breit-Wigner mass and width. Therefore, it is more sensible to use pole mass and width for calculations in either line shapes.

\section{Conclusions \label{sec:conclusions}}
In this work we have examined  the quasi-two-body $D\to SP$ decays and the three-body $D$ decays proceeding through intermediate scalar resonances. Our main results are:

\begin{itemize}
\item
In the $D\to SP_3\to P_1P_2P_3$ decays,  we cannot extract the two-body branching fractions $\B(D\to SP)$ for $S=f_0(980)$ and $a_0(980)$ due to the lack of information of $\B(S\to P_1P_2)$ (except for $a_0(980)\to\pi\eta$). For $S=\kappa/K_0^*(700)$ and $\sigma/f_0(500)$, the extracted two-body branching fractions are subject to large finite-width effects owing to their broad widths. Hence, for light scalars it is more sensible to study  $\B(D\to SP\to  P_1P_2P)$  directly and compare with experiment.

\item
We have considered the two-quark (scheme I) and four-quark (scheme II) descriptions of the light scalar mesons with masses below or close to 1 GeV.
Recent BESIII measurements of semileptonic charm decays favor the SU(3) nonet tetraquark description of the $f_0(500)$, $f_0(980)$ and $a_0(980)$ produced in charmed meson decay.
In Table \ref{tab:DtoSP:theory} we have calculated  $D\to SP_3\to P_1P_2P_3$ in schemes I and II. It is evident that scheme II agrees better with experiment for decays such as $D^+\to f_0\pi^+$ followed by $f_0\to \pi^+\pi^-$ and $D^+\to f_0K^+$ followed by $f_0\to \pi^+\pi^-$ or $f_0\to K^+K^-$. This again favors the tetraquark structure for light scalars. The predicted rates for $D^0\to f_0 P, a_0 P$ are generally smaller than experimental data by one order of magnitude, presumably implying the importance of $W$-exchange.

\item
The three-body decay modes $D^+\to \bar \kappa^0 (\to K_S\pi^0) \pi^+$, $D^+\to \ov K_0^* (\to K^-\pi^+) \pi^+$ and $D^+\to \ov K_0^* (\to K_S\pi^0) \pi^+$ are ideal for testing the validity of the factorization approach as they are free of $W$-annihilation contributions. $T$ and $C'$ amplitudes contribute constructively, contrary to the Cabibbo-allowed $D^+\to \ov K^0\pi^+$ decay where the interference between external and internal $W$-emission is destructive.

\item
Denoting the primed amplitudes $T'$ and $C'$ for the case when the emitted meson is a scalar meson, it is na{\"i}vely expected that $T'=C'=0$ for the neutral scalars $\sigma, f_0$ and $a_0^0$, $|T'|\ll|T|$ and $|C'|\ll|C|$ for the charged $a_0$ and $|T'|<|T|$ and $|C'|<|C|$ for the $\kappa$ and $K_0^*(1430)$. Beyond the factorization approximation, contributions proportional to the scalar decay constant $\bar f_S$ can be produced from vertex and hard
spectator-scattering corrections for the above-mentioned neutral scalars.

\item
We have studied the flavor operators $a_{1,2}(M_1M_2)$ for $M_1M_2=SP$ and $PS$ within the framework of QCD factorization.  Notice that $a_i(PS)$ and $a_i(SP)$ are very different as the former does not receive factorizable contributions.
While $a_{1,2}(SP)$ are similar for any light and heavy scalar mesons, $a_1(PS)$ and $a_2(PS)$ vary from neutral to the charged ones as shown in Table \ref{tab:aiPS}. The flavor operators $a_{1,2}(\pi a_0^\pm)$ are much greater than $a_{1,2}(\pi a_0^0)$.  In general,  $a_{1,2}(PS)$ become larger when the vector decay constants become smaller.

\item
For $f_0(980)$ and $a_0(980)$, we use the Flatt\'e line shape to describe both of them to take into account the threshold and coupled channel effects. For the very broad $\sigma/f_0(500)$ , we follow LHCb to employ a simple pole description.

\item
The annihilation amplitude inferred from the measurement of $D_s^+\to a_0^{+,0}\pi^{0,+}\to \eta\pi^{+,0}\pi^{0,+}$ is given by $|A|=(0.91\pm0.12)\times 10^{-6}\,{\rm GeV}$.  It is very sizable in the $SP$ sector, $|A/T|_{SP}\sim 1/2$, contrary to its suppression in the $P\!P$ sector with $|A/T|_{PP}\sim 0.18$.

\item
Since $\sigma$ and $\kappa$ are very broad, we have considered their finite-width effects characterized by the parameter $\eta_S$, whose deviation from unity measures the degree of departure from the NWA when the resonance width is finite. We find $\eta_{\sigma}$ and $\eta_{\kappa}$ to be of order $1.85 - 1.87$. The experimental branching fractions $\B(D^+\to\sigma\pi^+)$ and $\B(D^+\to\bar\kappa^0 \pi^+)$ should then read $(3.8\pm0.3)\times 10^{-3}$ and $(6.7^{+5.6}_{-4.5})\%$, respectively.

\item
For each scalar nonet (lighter and heavier one) we have 15 unknown parameters for the 8 topological amplitudes $T,C,E,A$ and $T',C',E',A'$. However, there are only 14 independent data to fit.  Moreover, since we need to introduce appropriate energy-dependent line shapes for the scalar mesons, it is not conceivable to extract the topological amplitudes from three-body decays as the decay rates cannot be factorized into the topological amplitude squared and the phase space factor.

\end{itemize}

\section*{Acknowledgments}

This research was supported in part by the Ministry of Science and Technology of R.O.C. under Grant Nos.~MOST-107-2119-M-001-034, MOST-110-2112-M-001-025 and MOST-108-2112-M-002-005-MY3, the National Natural Science Foundation of China under Grant No. 11347030, the Program of Science and Technology Innovation Talents in Universities of Henan Province 14HASTIT037.

\vskip 2.5cm

\end{document}